\documentclass[12pt]{article}

\usepackage[utf8]{inputenc}
\usepackage[T1]{fontenc}

\usepackage{setspace} 
\usepackage[vmargin = 1.25in, hmargin =1.25in]{geometry}

\usepackage[usenames,dvipsnames,svgnames]{xcolor}
\usepackage{amsmath, amssymb, amsfonts, graphicx, tikz,pdflscape, mathtools, amsthm, upgreek, bm,pgfplots,csvsimple,multirow, multicol, booktabs}
\usepgfplotslibrary{fillbetween}
\usepackage{verbatim}
\usepackage{natbib}
\usepackage{accents}
\newcommand{\ubar}[1]{\underaccent{\bar}{#1}}
\usepackage{needspace}
\def\citeapos#1{\citeauthor{#1}'s (\citeyear{#1})}
\usepackage{comment}
\usepackage[inline]{enumitem}
\usepackage{tocvsec2}
\usepackage{threeparttable}
\usetikzlibrary{calc, cd, arrows}
\usepackage[skip = 10pt]{caption}
\allowdisplaybreaks[1]

\definecolor{Blue}{RGB}{86,180,233}
\definecolor{Orange}{RGB}{230,159,0}
\definecolor{Green}{RGB}{0,158,115}
\definecolor{GmailBlue}{RGB}{42, 93, 176} 
\usepackage[
	pagebackref,
	colorlinks=true,
	citecolor= GmailBlue,
	linkcolor=GmailBlue,
	urlcolor = GmailBlue
]{hyperref}

\newcommand{\bibtexorder}[1]{}

\usepackage{pgfplots}
\usepgfplotslibrary{groupplots,colorbrewer}
\pgfplotsset{compat=newest}
\pgfplotsset{cycle list/Set1}
\usepackage{tikz}
\usetikzlibrary{matrix,calc,shapes,arrows.meta,positioning,math}
\tikzset{
    vertex/.style = {shape=circle,draw, minimum size = 1.8em, inner sep = 0pt},
    edge/.style = {->,> = latex}
}
\usetikzlibrary{shapes.misc}
\tikzset{cross/.style={cross out, draw=black, minimum size=2*(#1-\pgflinewidth), inner sep=0pt, outer sep=0pt},
cross/.default={1pt}}

\usepackage[capitalize,noabbrev]{cleveref}

\newtheoremstyle{break}
{}
{}
{\itshape}
{}
{\bfseries}
{}
{\newline}
{}

\theoremstyle{break}

\newtheorem*{theorem*}{Theorem}
\newtheorem*{cor*}{Corollary}

\newtheorem{prop}{Proposition}
\newtheorem{lem}{Lemma}

\crefname{prop}{Proposition}{Propositions}
\crefname{thm}{Theorem}{Theorems}
\crefname{lem}{Lemma}{Lemmas}
\crefname{blem}{Lemma}{Lemmas}

\theoremstyle{definition}
\newtheorem{defn}{Definition}

\newtheorem{rem}{Remark}
\newtheorem*{rem*}{Remark}

\newtheorem*{claim*}{Claim}

\def\a{\alpha}
\def\b{\beta}
\def\g{\gamma}
\def\d{\delta}
\def\e{\varepsilon}
\def\z{\zeta}

\def\th{\theta}

\def\k{\kappa}
\def\l{\lambda}

\def\s{\sigma}

\def\w{\omega}

\def\G{\Gamma}
\def\D{\Delta}
\def\Th{\Theta}

\def\W{\Omega}

\def\R{\mathbf{R}}

\def\AA{\mathcal{A}}
\def\BB{\mathcal{B}}

\def\FF{\mathcal{F}}

\def\KK{\mathcal{K}}

\def\MM{\mathcal{M}}

\def\PP{\mathcal{P}}

\def\SS{\mathcal{S}}

\def\UU{\mathcal{U}}

\def\YY{\mathcal{Y}}

\def\E{\mathbf{E}}

\DeclareMathOperator{\supp}{supp} 
\DeclareMathOperator*{\argmax}{argmax}
\DeclareMathOperator*{\maz}{maximize}

\DeclareMathOperator{\conv}{conv}
\DeclareMathOperator{\cconv}{\overline{conv}}

\DeclareMathOperator{\aff}{aff}

\DeclareMathOperator{\marg}{marg} 

\DeclareMathOperator{\lsc}{lsc}

\newcommand{\ang}[1]{\langle #1 \rangle} 

\newcommand{\Ang}[1]{\left\langle #1 \right\rangle}

\newcommand{\paren}[1]{( #1 )} 

\newcommand{\Paren}[1]{\left( #1 \right)}

\newcommand{\Brac}[1]{\left[ #1 \right]}

\newcommand{\set}[1]{\{ #1 \}} 

\newcommand{\Set}[1]{\left\{ #1 \right\}}

\newcommand{\de}{\mathop{}\!\mathrm{d}}

\usepackage{datetime}
\newdateformat{specialdate}{\THEDAY~\monthname[\THEMONTH] \THEYEAR}

\title{Robust Robustness\thanks{For useful feedback, we thank audiences at SAET, the Penn Mini-Conference on Economic Theory, NASMES, MIT/Harvard, Yale, University of Chicago, Arizona State, the EAYE Annual Meeting, and the BSE Choice and Decision Workshop. For helpful discussions, we thank  Ben Brooks, Eddie Dekel, Jeff Ely, Piotr Dworczak, Joel Flynn, Drew Fudenberg, Yingni Guo, Lars Peter Hansen, Tibor Heumann, Alex Jakobsen, Philippe Jehiel, Peter Klibanoff, Elliot Lipnowski, Krist\'{o}f Madar\'{a}sz, George Mailath, Suraj Malladi, Stephen Morris, Pietro Ortoleva, Alessandro Pavan, Harry Pei, Jacopo Perego, Doron Ravid, Phil Reny, Larry Samuelson, Marciano Siniscalchi, Rani Spiegler, Lorenzo Maria Stanca, Juuso Toikka, Juuso V\"{a}lim\"{a}ki, Alex Wolitzky, and Kai Hao Yang.  This paper was initially written while Ian Ball was visiting Northwestern University; we are grateful for their hospitality.}}

\date{\specialdate\today}

\author{Ian Ball \and Deniz Kattwinkel}

\begin{document}

\begin{titlepage} 

\maketitle \thispagestyle{empty} 

\begin{abstract} 
We propose a refinement of the maxmin approach to robustness. A mechanism's payoff guarantee over an ambiguity set is \emph{robust} if the guarantee is approximately satisfied at priors near the ambiguity set (in the weak topology). We show that many maxmin-optimal mechanisms in the literature give  payoff guarantees that are not robust. Such mechanisms are often tailored to degenerate worst-case priors, making them simple but fragile. Conversely, some commonly used ambiguity sets satisfy a structural property, termed richness, ensuring that every associated payoff guarantee is robust. We show how to slightly enlarge any ambiguity set to make it rich. 
\end{abstract}

\noindent Keywords: robust mechanism design; maxmin expected utility; simple mechanisms

\end{titlepage}

\section{Introduction}

The standard Bayesian approach to mechanism design assumes that the designer has a prior over the relevant set of states. In practice, the designer may find it difficult to quantify her uncertainty with an exact prior. To capture this, the maxmin approach models the designer's uncertainty as a set of priors called the \emph{ambiguity set}. The designer evaluates each mechanism according to its \emph{payoff guarantee}, i.e., its worst-case expected payoff over the priors in the ambiguity set. But just as it is difficult for the designer to form an exact prior, it may be difficult for the designer to form an exact ambiguity set. This raises the concern that a maxmin-optimal mechanism may perform much worse than its payoff guarantee at priors slightly outside the ambiguity set. 

This fragility can be illustrated in the following simple example. Consider the standard monopoly selling problem. The designer (seller) has a single good. She is uncertain of the buyer's valuation, but she is unable to quantify her uncertainty with a single prior. She knows only that the valuation distribution has median $\l$, for some fixed $\l > 0$. She evaluates each implementable social choice function according to its revenue guarantee over all valuation distributions with median $\l$. It can be verified that the best possible revenue guarantee is $\l/2$. This guarantee is uniquely achieved by posting a price of $\l$, so that the agent buys if and only if his valuation is at least $\l$. But the revenue guarantee from this posted price is fragile. For any $\e \in (0,\l)$, there exists a valuation distribution with median $\l - \e$ under which this posted price yields revenue $0$ (because the agent never buys). 

In order to analyze the robustness of the maxmin approach, we consider a general Anscombe--Aumann setting, enriched with a topology on the state space. The state represents all aspects of the environment that are unknown to the designer, such as agents' preferences or technology.  The topology on the state space reflects which states the designer finds difficult to distinguish when she forms her beliefs. The space of priors is endowed with the weak topology induced by the topology on the state space. The designer's uncertainty about the state is represented by a set of priors called the ambiguity set. The designer evaluates each social choice function according to its payoff guarantee over all priors in the ambiguity set. Given a feasible set of social choice functions, the designer chooses a social choice function with the greatest payoff guarantee. For example,  in the monopoly selling problem, the feasible set contains all social choice functions that satisfy incentive compatibility and participation constraints. 

In this general setting, we introduce a new robustness criterion: the payoff guarantee from a social choice function over an ambiguity set is \emph{robust} if the expected payoff from the social choice function approximately satisfies the guarantee at all priors sufficiently near the ambiguity set. That is, the guarantee is \emph{not} robust if the expected payoff falls below the guarantee by at least a fixed amount at some priors arbitrarily close to the ambiguity set. Thus, a non-robust payoff guarantee is useful only if the designer has confidence in the boundary of the ambiguity set to arbitrary precision.
 
We apply our robustness criterion to prominent maxmin-optimal mechanisms proposed in the literature in various settings: monopoly screening (with revenue and regret objectives), persuasion, delegation, and moral hazard. These mechanisms often take ``simple'' forms. In many of these examples, however, the payoff guarantee from the proposed maxmin-optimal mechanism is not robust in our sense. In some cases, the expected payoff from the optimal mechanism drops to its lowest possible value at some priors arbitrarily close to the ambiguity set. 

The simple maxmin-optimal mechanisms identified in the literature are often interpreted as evidence that mechanisms which perform well over a range of priors tend to be simple. Our analysis challenges this interpretation. To solve for a maxmin-optimal mechanism, the standard approach is to identify a saddle point of the designer's zero-sum game against Nature. In this game, the designer chooses a feasible social choice function, and Nature chooses a prior in the ambiguity set. For any finite-support prior, there is typically an associated Bayesian optimal mechanism that is ``simple.'' Any such pair constitutes a saddle point for a range of nonsingleton ambiguity sets (\cref{res:Bayesian_robustness}). For any of these ambiguity sets, the simple mechanism in the saddle point is maxmin-optimal. This simplicity, however, reflects the simplicity of the Bayesian solution at a particular finite-support prior, not a general consequence of the maxmin approach. 

Moreover, a mechanism obtained from a saddle point with a finite-support prior typically yields a non-robust payoff guarantee. Such a mechanism is precisely tailored to the states where the point masses are located. If these point masses are slightly perturbed, then the designer's expected payoff from the mechanism drops dramatically. \cref{res:saddle} formalizes this fragility in a profit-maximization problem. Given any saddle point with a finite-support prior, the associated optimal mechanism's profit guarantee, if strictly positive, is non-robust. 

Having analyzed prominent failures of robustness, we turn to positive results. First, we show how robustness can be assured by the structure of the ambiguity set. An ambiguity set is \emph{rich} if, for any prior sufficiently near the set, it is possible to reallocate a small amount of probability across states so as to obtain a new prior inside the ambiguity set. \cref{res:rich} says that every rich ambiguity set is \emph{uniformly robust}, which implies that every social choice function's payoff guarantee over that ambiguity set is robust. Therefore, with a rich ambiguity set, the analyst is assured, before she computes a maxmin-optimal mechanism, that the payoff guarantee from that mechanism will be robust. \cref{res:robust_sets} identifies two commonly used classes of ambiguity sets that are rich: \emph{continuous moment sets}, which are defined by restrictions on the expectations of continuous functions of the state; and \emph{metric neighborhoods},  which contain all priors within a fixed distance of a given set, where the distance is measured using the Wasserstein metric or any other convex metric that induces the weak topology.

When solving for a maxmin-optimal mechanism, we propose that the analyst check whether the associated payoff guarantee is robust. If the guarantee is not robust, then there are various ways to proceed. The following algorithm recovers robustness by enriching the original ambiguity set in the maxmin objective. On the space of priors, choose a convex metric that induces the weak topology. Consider a new ambiguity set that contains all priors within a fixed distance of the original ambiguity set. Compute a maxmin-optimal mechanism with respect to this new ambiguity set.  The payoff guarantee from this new mechanism over the new ambiguity set is robust because metric neighborhoods are uniformly robust (by \cref{res:rich,res:robust_sets}). 

In this algorithm, the radius of the metric neighborhood parametrizes the designer's concern for misspecification. Crucially, every mechanism's payoff guarantee is continuous as a function of this radius $r > 0$. Moreover, we show that the new mechanism maximizes a particular variational objective in which the ambiguity index is proportional to the distance from the original ambiguity set (\cref{res:maxmin_variational_equivalence}). We apply this algorithm to the regret-minimizing monopoly selling problem in \cite{BergemannSchlag2008}. Their proposed maxmin-optimal mechanism is a particular randomized posted price. For some parameters, this price distribution has an atom, in which case the associated payoff guarantee is not robust. By contrast, the mechanism obtained from our algorithm always randomizes continuously over prices. 

The rest of the paper is organized as follows. \cref{sec:model} introduces the setting and defines our notion of robustness. \cref{sec:fragile} applies our robustness criterion to maxmin-optimal mechanisms proposed in the literature. In \cref{sec:saddle}, we diagnose the failure of robustness in the examples by illustrating how saddle points with finite-support priors yield non-robust payoff guarantees. \cref{sec:classification} analyzes uniform robustness and presents an algorithm for constructing a maxmin-optimal mechanism with a robust payoff guarantee. \cref{sec:literature} discusses related literature. \cref{sec:conclusion} is the conclusion. The main proofs are in \cref{sec:main_proofs}. The online appendix contains additional results (\cref{sec:alternative_axiomatizations}) and technical proofs (\cref{sec:technical_proofs}).

\section{Model} \label{sec:model}

\subsection{Maxmin setting}
 
There is a state space $\Th$, which is a Polish topological space endowed with its Borel $\s$-algebra. The state represents all aspects of the environment that are unknown to the designer, such as agents' preferences or technology. There is a decision space $X$, which is endowed with a $\s$-algebra. Denote by $\D (\Th)$ (respectively, $\D(X)$) the space of probability measures on $\Th$ (respectively, $X$). A social choice function is a measurable function $f \colon \Th \to \D (X)$. The designer has a bounded, measurable utility function $u \colon X \times \Th \to \R$, which we extend linearly to $\D(X) \times \Th$.\footnote{In some of our examples, the function $u$ will not be bounded, but utility will be bounded over the relevant outcomes.} The designer's state-dependent utility can capture different objectives, including the minimization of ex-post regret.

This setting is the same as the classical Anscombe--Aumann framework, except that we also consider a topology on the state space.\footnote{We use the terminology of mechanism design because mechanism design is our primary motivation. In the language of decision theory, the designer is the decision-maker; decisions are consequences; and social choice functions are Anscombe--Aumann acts.} This topology reflects which states the designer
finds difficult to distinguish when she is forming her beliefs. That is, the designer cannot confidently choose between two priors if the only difference between them is that probability is moved between nearby states, e.g., from a valuation of $\$$10.23 to $\$$10.24 in the monopoly selling example. The topology is important for our notion of robustness below. 

The designer evaluates each social choice function $f$ according to the objective 
\begin{equation} \label{eq:objective}
    \inf_{\pi \in \Pi} \int_{\Th} u(f(\th), \theta) \de \pi (\th),
\end{equation}
where $\Pi$ is a nonempty subset of $\D(\Th)$ called the \emph{ambiguity set}. We take an infimum rather than a minimum because we have not made assumptions on $\Pi$ to guarantee the existence of a minimizer.\footnote{\label{ft:GS}If $u$ is state-independent, then this objective also has a
\cite{GilboaSchmeidler1989} representation as a minimum over a set of finitely additive probability measures; see \cref{sec:finitely_additive} for details. With state-dependent utility, maxmin expected utility is axiomatized by \cite{Hill2019} in a setting with a finite state space.} The expression in \eqref{eq:objective} is the designer's \emph{payoff guarantee from $f$ over $\Pi$}, i.e., the designer's worst-case expected payoff from the social choice function $f$ over all priors in the ambiguity set $\Pi$. The payoff guarantee depends on the social choice function $f$ only through the induced \emph{value function} $u \circ f$ defined by $(u \circ f)(\th) = u(f(\th), \th)$. Hereafter, we use inner product notation for integrals, so the integral in \eqref{eq:objective} will be denoted by $\ang{ u \circ f, \pi}$. 

Given a feasible set $\FF$ of social choice functions, the designer's problem is 
\begin{equation} \label{eq:designer_problem}
    \maz_{f \in \FF} \Brac{ \inf_{\pi \in \Pi} \ang{u \circ f, \pi}}.
\end{equation}
The term in brackets is the payoff guarantee in \eqref{eq:objective}, expressed in our new notation. A social choice function $\hat{f}$ that solves \eqref{eq:designer_problem} is \emph{maxmin optimal} (with respect to the feasible set $\FF$ and the ambiguity set $\Pi$).

In our motivating mechanism-design applications, the feasible set $\FF$ will contain all implementable social choice functions. For example, in a screening problem, the state $\th$ is the agent's type, and the set $\FF$ contains all social choice functions that satisfy incentive compatibility and participation constraints. We call a mechanism \emph{maxmin optimal} if it implements a maxmin-optimal social choice function. We will also speak of a mechanism's payoff guarantee when the induced social choice function is understood. 

\subsection{Robustness of a payoff guarantee}

We now formally define our notion of robustness. Endow the space $\D(\Th)$ with the weak topology associated with the topology on the state space $\Th$.\footnote{That is, the weak topology induced by integration against bounded, continuous functions on $\Th$. A sequence $(\pi_n)$ \emph{weakly converges} to $\pi$ if $\ang{ h, \pi_n} \to \ang{h, \pi}$ for each bounded, continuous function $h \colon \Th \to \R$.} The weak topology reflects the underlying topology on the state space; for example,  if the sequence $(\th_n)$ in $\Th$ converges to $\th$, then the associated sequence $(\d_{\th_n})$ of unit masses in $\D(\Th)$ weakly converges to $\d_{\th}$, even though distinct unit masses have disjoint supports. Unless otherwise indicated, topological statements about $\D(\Th)$ are with respect to the weak topology. 

\begin{defn}[Robustness] \label{def:robustness} Let $\Pi$ be a nonempty subset of $\D (\Th)$. Let $f \colon \Th \to  \D (X)$ be a social choice function. The \emph{payoff guarantee from $f$ over $\Pi$ is robust} if for every sequence $(\pi_n)$ in $\D(\Th)$ that converges to a prior in the closure of $\Pi$,
     \begin{equation} \label{eq:robust}
        \liminf_{n} \ang{u \circ f, \pi_n} \geq \inf_{\pi \in \Pi} \ang{ u \circ f, \pi}.
    \end{equation}
\end{defn}

In words, the payoff guarantee from  $f$ over $\Pi$ is robust if the expected payoff from $f$ approximately satisfies the guarantee at all priors
sufficiently near $\Pi$. The payoff guarantee from $f$ over $\Pi$ is \emph{not} robust if and only if there exists a constant $\a > 0$ and a sequence $(\pi_n)$ converging to a prior in the closure of $\Pi$ such that for all $n$,
\begin{equation} \label{eq:not_robust}
    \ang{ u \circ f, \pi_n} \leq \inf_{\pi \in \Pi} \ang{u \circ f, \pi} - \a.
\end{equation}
Inequality \eqref{eq:not_robust} can hold only if $\pi_n$ is outside $\Pi$. If $\Pi = \D(\Th)$, then no priors are outside $\Pi$. Therefore, every social choice function's payoff guarantee over $\D(\Th)$ is robust.  Recall our standing assumption that $u$ is bounded. In some applications, $u$ will be unbounded, but we can safely apply \cref{def:robustness} to any social choice function $f$ for which $u \circ f$ is bounded. 

Our definition of robustness can be applied to the payoff guarantee from any social choice function $f$ over any ambiguity set $\Pi$. In a given design problem, we propose checking whether the payoff guarantee from the proposed maxmin-optimal social choice function (over the specified ambiguity set) is robust.

Using the weak topology on $\D(\Th)$ is what gives our robustness criterion bite. By contrast,  in \citeapos{GilboaSchmeidler1989} axiomatization of maxmin preferences, there is no topology on the state space, so the space of priors inherits a stronger topology under which few priors are ``near'' each ambiguity set.\footnote{\cite{GilboaSchmeidler1989} consider the space of finitely additive probability measures endowed with the topology of pointwise convergence on measurable sets.} With respect to that stronger topology on $\D(\Th)$, the map $\ang{ u \circ f, \cdot}$ on $\D(\Th)$ is continuous for any bounded value function $u \circ f$. If we used that stronger topology on $\D(\Th)$ in our definition of robustness, then our robustness criterion would hold for every social choice function $f$ and every ambiguity set $\Pi$.

\begin{rem}[Robustness in discrete models] \label{rem:discrete} If the state space $\Th$ is endowed with the discrete topology, then robustness has no bite because for each social choice function $f$, the map $\ang{ u \circ f, \cdot}$ on $\D(\Th)$ is continuous. In  many applications with discrete states, however, it is natural to think of these states as being embedded in a continuous state space. For example, consider a model in which a buyer's valuation equals $\th_H$ with probability $p$ and 
equals $\th_L$ with probability $1-p$. If we take $\Th = \{ \th_L, \th_H \}$, then robustness has no bite because it only considers perturbations of the \emph{probabilities} on $\th_L$ and $\th_H$. But if we take $\Th$ to be an interval containing $\th_L$ and $\th_H$ in its interior, then robustness does have bite because it also takes into account perturbations of the \emph{valuations} $\th_L$ and $\th_H$.
\end{rem}

The interpretation of our robustness criterion hinges on the topology on $\Th$ and the associated weak topology on $\D(\Th)$. In the weak topology, two kinds of perturbations are considered small---reallocating a \emph{small} probability between arbitrary states and reallocating any amount of probability between \emph{nearby} states. Therefore, our robustness criterion is natural if the designer, when forming her beliefs, is unable to confidently choose between priors distinguished by such small perturbations. In this case, it is difficult to imagine how the designer could be reassured by a payoff guarantee that is substantially violated just outside the ambiguity set. Because the maxmin representation makes a binary distinction between priors inside and outside the ambiguity set, the exact boundary is often arbitrary. Of course, the designer could consider a larger ambiguity set, but unless the ambiguity set is enlarged to the full space $\D(\Th)$ (in which case the optimal payoff guarantee is often trivial), there will still be priors just outside this new ambiguity set. Alternatively, the designer could abandon maxmin and instead consider a variational objective \citep{MaccheroniMarinacciRustichini2006}, which assigns different ``penalties'' to different priors; this approach is discussed further in \cref{sec:algorithm}.

Our definition of robustness considers perturbations of the priors in the ambiguity set $\Pi$. Robustness can be characterized in terms of perturbations of the ambiguity set $\Pi$ itself, provided that $\Pi$ is compact. Let $\mathcal{K}$ denote the collection of nonempty, compact subsets of $\D( \Th)$. Endow $\KK$ with the topology induced by the Hausdorff metric.\footnote{\label{ft:metrization}For all metrizations of the weak topology on $\D(\Th)$, the associated Hausdorff metric induces the same topology on $\KK$ \citep[3.91, p.~120]{AliprantisBorder2006}. Note that our definition of robustness depends only on the weak topology on $\D(\Th)$, not its metrization.}

\begin{prop}[Robustness as lower semicontinuity in the ambiguity set] \label{res:Hausdorff_lsc} 
Let $\Pi$ be in $\KK$. Let $f \colon \Th \to \D(X)$ be a social choice function. The payoff guarantee from $f$ over $\Pi$ is robust if and only if the map $ \Pi' \mapsto \inf_{\pi \in \Pi'} \ang{u \circ f, \pi}$ on $\mathcal{K}$ is lower semicontinuous at $\Pi$.
\end{prop}

Intuitively, robustness demands that the payoff guarantee from $f$ over $\Pi$ does not drop discontinuously when the ambiguity set $\Pi$ is perturbed.

\section{Robustness in maxmin applications} \label{sec:fragile}

In this section, we apply our robustness criterion to prominent maxmin-optimal mechanisms proposed in the literature in various settings: monopoly screening (with revenue and regret objectives), persuasion, delegation, and moral hazard. In many of these examples, we show that the payoff guarantee from the proposed maxmin-optimal mechanism is not robust.  To guide our discussion, we first relate robustness to the continuity of the induced value function. 

\subsection{Continuous value functions and robustness}

Before turning to the applications, we give a preliminary result on the relationship between robustness and the continuity of the induced value function. For any function $v \colon \Th \to \R$, let $D_-(v)$ denote the set of points $\th$ in $\Th$ at which $v$ is not lower semicontinuous. 

\begin{prop}[Robustness and continuity] \label{res:robustness_continuity} Let $f \colon \Th \to \D(X)$ be a social choice function. 
\begin{enumerate}[label = (\roman*), ref = \roman*]
    \item \label{it:both}  The payoff guarantee from $f$ over $\Pi$ is robust for every ambiguity set $\Pi$ if and only if $u \circ f$ is continuous. 
    \item \label{it:closed}  The payoff guarantee from $f$ over $\Pi$ is robust for every closed ambiguity set $\Pi$ if and only if $u \circ f$ is lower semicontinuous. 
    \item \label{it:lsc} Given a prior $\pi$ in $\D(\Th)$, the payoff guarantee from $f$ over $\{ \pi \}$ is robust if and only if $\pi ( D_-( u \circ f)) = 0$.
    \item \label{it:closure} Given a closed ambiguity set $\Pi$, the payoff guarantee from $f$ over $\Pi$ is robust if $\pi ( D_- ( u \circ f)) = 0$ for every prior $\pi$ in $\Pi$. 
\end{enumerate}
\end{prop}

Parts \ref{it:both} and \ref{it:closed} consider robustness with respect to every ambiguity set in some class.  By the portmanteau theorem, the map $\ang{u \circ f, \cdot}$ on $\D(\Th)$ inherits  continuity (lower semicontinuity) from the function $u \circ f$ on $\Th$. If $\ang{ u \circ f, \cdot}$ is continuous, then perturbing any prior in the ambiguity set has a small effect on the expectation of $u \circ f$. Hence, the payoff guarantee from $f$ over any ambiguity set is robust. If $\ang{u \circ f, \cdot}$ is lower semicontinuous, then the expectation of $u \circ f$ can jump down, but not up, at the limit of a sequence of priors. Thus, the robustness inequality \eqref{eq:robust} is satisfied as long as the limiting prior is inside the ambiguity set, which holds whenever the ambiguity set is closed.  

Parts \ref{it:lsc} and \ref{it:closure} consider robustness with respect to a fixed ambiguity set.  If the ambiguity set contains a single prior, then the payoff guarantee reduces to the Bayesian expected payoff under that prior. In this case, part~\ref{it:lsc} gives a characterization: the Bayesian expected payoff from $f$ under prior $\pi$ is robust if and only if the induced value function $u \circ f$ is lower semicontinuous $\pi$-almost surely. For any closed ambiguity set $\Pi$, part~\ref{it:closure} gives a sufficient condition for robustness: the payoff guarantee from $f$ over $\Pi$ is robust if, for each $\pi$ in $\Pi$, the induced value function $u \circ f$ is lower semicontinuous $\pi$-almost surely. This sufficient condition is not necessary, as can be seen from \cref{res:rich,res:robust_sets} below.

\cref{res:robustness_continuity} can be used to confirm that the payoff guarantee from a particular mechanism is robust. It suffices to check that the induced value function is continuous. But many simple mechanisms of interest, such as posted prices, induce \emph{discontinuous} value functions. For these mechanisms, the robustness of the payoff guarantee depends on the structure of the ambiguity set.

\begin{rem}[Absolutely continuous priors] \label{res:AC} Suppose that $\Th$ is a subset of Euclidean space. \cref{res:robustness_continuity} has powerful implications for priors that are absolutely continuous (with respect to Lebesgue measure). Let $\Pi$ be a closed ambiguity set that contains only absolutely continuous priors. By \cref{res:robustness_continuity}.\ref{it:closure},  the payoff guarantee from a social choice function $f$ over $\Pi$ is robust as long as the induced value function $u \circ f$ is continuous Lebesgue-almost everywhere. In the monopoly selling problem and the multi-good monopoly problem (under both revenue and regret objectives), for each incentive-compatible social choice function $f$, the value function $u \circ f$ is continuous Lebesgue-almost everywhere; for details, see \cref{sec:AC_proof}. So in these applications, if $\Pi$ is a closed set of absolutely continuous priors, then every mechanism's payoff guarantee over $\Pi$ is robust. 
\end{rem}

\subsection{Monopoly selling: revenue maximization}
\label{sec:monopoly}

\begin{figure}
\centering
\begin{tikzpicture}
    \begin{axis}[
        width = 0.5 \textwidth,
        height = 0.35 \textwidth,
        xmin = 0, xmax = 1,
        ymin = 0, ymax = 0.7,
        axis lines = middle,
        xtick = {0.4},
        xticklabels = {$\l$},
        ytick =  {0.4},
        yticklabels = {$\l$},
        clip = false, 
        xlabel = {$\th$},
        xlabel style={at=(current axis.right of origin), anchor=west},
    ]
        \addplot[very thick, Blue] coordinates {(0.4,0.4) (1,0.4)} node[pos=0.8, above] {$u \circ \hat{f}$};
        \addplot[very thick, Blue] coordinates {(0,0) (0.4,0)};
        \addplot[dashed, Blue] (0.4,0) -- (0.4,0.4);
        \filldraw[Blue] (0.4,0.4) circle (2pt);
        \filldraw[Blue, fill=white] (0.4,0) circle (2pt);
    \end{axis}
\end{tikzpicture}
\quad
\begin{tikzpicture}
    \begin{axis}[
        width = 0.5 \textwidth,
        height = 0.35 \textwidth,
        xmin = 0, xmax = 3.5,
        ymin = 0, ymax = 1.6,
        axis lines = middle,
        xtick = {0.173755, 2.96448},
        xticklabels = {$\ubar{\th}$,  $\bar{\th}$},
        ytick = {0.173755},
        yticklabels = {$\ubar{\th}$},
        clip = false, 
        xlabel = {$\th$},
        xlabel style={at=(current axis.right of origin), anchor=west},
    ]
        \addplot[very thick, Blue] coordinates {(0,0) (0.17,0)};
        \addplot[very thick, Blue, smooth, domain=0.173755:2.96448] { -0.0889846 + 0.527587*x - 0.0889846*x^2 } node[pos=1.05, above] {$u \circ \hat{f}$};
        \addplot[very thick, Blue] coordinates {(2.96448,0.693027) (3.5,0.693027)};
    \end{axis}
\end{tikzpicture}\\ 
\vspace{5pt}
\begin{tikzpicture}
    \begin{axis}[
         width = 0.5 \textwidth,
        height = 0.35 \textwidth,
        xmin = 0, xmax = 1,
        ymin = 0, ymax = 1.1,
        axis lines = middle,
        xtick = {0.4},
        xticklabels = {$\l$ \vphantom{$\b$}},
        ytick =  {0.5,1},
        yticklabels = {$\frac{1}{2}$,$1$},
        clip = false, 
        xlabel = {$\th$},
        xlabel style={at=(current axis.right of origin), anchor=west},
    ]
        \addplot[very thick, Orange] coordinates {(0.4,1) (1,1)} node[pos=0.8, above] {$F_{\hat{\pi}}$};
        \addplot[very thick, Orange] coordinates {(0,0.5) (0.4,0.5)};
        \addplot[dashed, Orange] (0.4,0.5) -- (0.4,1);
        \addplot[dashed, Orange] (0,0) -- (0,0.5);
        \filldraw[Orange] (0.4,1) circle (2pt);
        \filldraw[Orange, fill=white] (0.4,0.5) circle (2pt);
        \filldraw[Orange] (0,0.5) circle (2pt);
        \filldraw[Orange, fill=white] (0,0) circle (2pt);
    \end{axis}
\end{tikzpicture}
\quad
\begin{tikzpicture}
    \begin{axis}[
        width = 0.5 \textwidth,
        height = 0.35 \textwidth,
        xmin = 0, xmax = 3.5,
        ymin = 0, ymax = 1.1,
        axis lines = middle,
        xtick = {0.17, 2.96448},
        xticklabels = {$\ubar{\th}$,  $\bar{\th}$},
        ytick = {1},
        yticklabels = {1},
        clip = false, 
        xlabel = {$\th$},
        xlabel style={at=(current axis.right of origin), anchor=west},
    ]
        \addplot[very thick, Orange] coordinates {(0,0) (0.173755,0)};
        \addplot[very thick, Orange, smooth, domain=0.173755:2.96448] { 1-0.173755/x };
        \addplot[very thick, Orange] coordinates {(2.96448,1) (3.5,1)}node[pos=0.5, above] {$F_{\hat \pi}$};
        \addplot[dashed, Orange] (2.96448,1-0.173755/2.96448) -- (2.96448,1);
         \filldraw[Orange] (2.96448,1) circle (2pt);
        \filldraw[Orange, fill=white] (2.96448,1-0.173755/2.96448) circle (2pt);
    \end{axis}
\end{tikzpicture}

\caption{Monopoly revenue maximization with median restriction (left column) and moment restrictions (right column).}
\label{fig:monopoly}
\end{figure}

Consider the standard monopoly selling problem. The seller has a single good. The state $\th \in \Th = \R_+$ is the buyer's valuation for the good. 

First, we formalize the example from the introduction.  The seller seeks a revenue guarantee over the ambiguity set $\Pi$ containing all priors that have $\l$ as a median. This revenue guarantee is uniquely maximized by posting the price $p  = \l$; the agent buys if and only if $\th \geq \l$.\footnote{To ensure that the designer's problem has a solution, we assume that ties are broken in the designer's favor, as is standard.} Denote this social choice function by $\hat{f}$. The induced value function $u \circ \hat{f}$ is shown in the top left panel of \cref{fig:monopoly}. The seller's payoff guarantee from $\hat{f}$ over $\Pi$ equals $\l/2$. This worst-case payoff is achieved at the prior $\hat{\pi} = \d_{0}/2 + \d_{\l}/2$. The associated cumulative distribution function, $F_{\hat{\pi}}$, is shown in the bottom left panel of \cref{fig:monopoly}. The payoff guarantee from $\hat{f}$ over $\Pi$ is not robust. For any $\e \in (0,\l)$, if the prior $\hat{\pi}$ is perturbed to $\pi_\e = \d_0/2 + \d_{\l - \e}/2$, then the seller's expected revenue from $\hat{f}$ drops to $0$ (because the good is never purchased).

In this example, the pair $(\hat{f}, \hat{\pi})$ is a saddle point of the zero-sum game between the designer and Nature. In this game, the designer chooses a social choice function $f$ from the feasible set $\FF$, and Nature simultaneously chooses a prior $\pi$ from the ambiguity set $\Pi$. Maxmin design problems are commonly solved by finding such a saddle point. The saddle point can provide insights about robustness. Here, the payoff guarantee from $\hat{f}$ over $\Pi$ is not robust because $\hat{f}$ is tailored to the state $\th = \l$ where $\hat{\pi}$ has a point mass; $\hat{f}$ performs well in the state $\th = \l$, but $\hat{f}$ yields zero revenue at arbitrarily nearby states, which are outside the support of $\hat{\pi}$. This logic is generalized in \cref{res:saddle} below.

\cite{CarrascoEtal2018} specify a different ambiguity set in the same revenue-maximization problem. In their leading case, the ambiguity set $\Pi$ contains all distributions over $\Th = \R_+$ with mean $\mu_1$ and second moment at most $\mu_2$, where the parameters $\mu_1$ and $\mu_2$ satisfy $\mu_2 > \mu_1^2 > 0$. \citet[Proposition 4, p.~258]{CarrascoEtal2018} solve for a saddle point $(\hat{f}, \hat{\pi})$ of the designer's game against Nature. The social choice function $\hat{f}$ is implemented by randomizing the price over some interval $[\ubar{\th}, \bar{\th}]$ according to a continuous density function.  Nature's optimal prior $\hat{\pi}$ has support $[\ubar{\th}, \bar{\th}]$. The value function $u \circ \hat{f}$ is continuous, so the payoff guarantee from $\hat{f}$ over $\Pi$ is robust (by \cref{res:robustness_continuity}.\ref{it:both}). The value function $u \circ \hat{f}$ and the cumulative distribution function $F_{\hat{\pi}}$ associated with $\hat{\pi}$ are shown in the right column of \cref{fig:monopoly}, for the case $\mu_1 = 2/3$ and $\mu_2 = 1$. In this case, Nature's distribution $\hat{\pi}$ contains a point mass at $\bar{\th}$, but the price distribution implementing $\hat{f}$ does not put mass on $\bar{\th}$. We show below (\cref{res:rich,res:robust_sets}) in a general setting that if the ambiguity set is defined by restrictions on continuous moments, then every mechanism's payoff guarantee is robust. 

\subsection{Monopoly selling: regret minimization} \label{sec:monopoly_selling_regret}

\begin{figure}
\centering
\begin{tikzpicture}
    \begin{axis}[
        xmin = 0, xmax = 1.2,
        ymin = 0, ymax = 0.6,
        axis lines = middle,
        xtick = {0.5, 1},
        xticklabels = {$\ubar{\th}$, $1$},
        ytick =  {0.346574, 0.5},
        yticklabels = {$R$, $\ubar{\th}$},
        clip = false, 
        xlabel = {$\th$},
        xlabel style={at=(current axis.right of origin), anchor=west},
    ]
        \addplot[very thick, Blue] coordinates {(0,0) (0.5,0.5)};
        \addplot[very thick, Blue] coordinates {(0.5, 0.346574) (1,0.346574)}node[pos=0.6, above] {$- u \circ \hat{f}$};
        \addplot[very thick, Blue, smooth, domain=1:1.2]{0.346574 + x - 1};
        \addplot[dashed, Blue] (0.5,0.346574) -- (0.5,0.5);
        \filldraw[Blue] (0.5,0.346574) circle (2pt);
        \filldraw[Blue, fill=white] (0.5,0.5) circle (2pt);
    \end{axis}
\end{tikzpicture}

\caption{Monopoly regret-minimization with support restriction}
\label{fig:regret}
\end{figure}

In the monopoly selling problem, an alternative objective for the monopolist is to minimize ex-post regret (i.e., forgone revenue relative to fully extracting the agent's realized valuation). Under a Bayesian prior, revenue maximization and regret minimization yield the same solution. With a worst-case objective, however, the solutions of these two problems can be quite different. 

In \cite{BergemannSchlag2008}, the designer evaluates each social choice function according to its worst-case regret over all valuations $\th$ in $[\ubar{\th}, 1]$, where the parameter $\ubar{\th}$ satisfies $0 \leq \ubar{\th} < 1$. Setting the upper endpoint to $1$ is simply a normalization of the valuation units. To represent this regret-minimization problem in our framework, let $\Th = [0, M]$ for some $M > 1$; this bound $M$ ensures that under every mechanism, the associated regret function is bounded. Let the designer's utility $u$ be \emph{negative} ex-post regret. Thus, $u$ equals revenue net of the agent's true valuation $\th \in \Th$. Let the ambiguity set $\Pi$ contain all valuation distributions that concentrate on $[\ubar{\th}, 1]$. \citet[Proposition 1, p.~565]{BergemannSchlag2008} solve for the unique maxmin-optimal social choice function $\hat{f}$. There are two cases. If $\ubar{\th} \leq 1/e$, then $\hat{f}$ is implemented by randomizing over prices $p$ in the interval $[1/e,1]$ with density $1/p$. This distribution has no point masses, so the induced value function is continuous and hence the regret guarantee is robust (by \cref{res:robustness_continuity}.\ref{it:both}). If $\ubar{\th} > 1/e$, then the optimal price distribution puts mass $1 + \ln \ubar{\th}$ on the price $p = \ubar{\th}$ and has density $1/p$ over $[\ubar{\th}, 1]$. \cref{fig:regret} plots the induced regret function, $- u \circ \hat{f}$, with $\ubar{\th} = 1/2$.  This price distribution ensures that over the interval $[\ubar{\th}, 1]$, revenue is increasing with unit slope, so regret is constant. The regret guarantee from $\hat{f}$ over $\Pi$ is $R = -\ubar{\th} \ln \ubar{\th}$, but this guarantee is not robust. For $\th \geq \ubar{\th}$, the agent buys with probability at least $1 + \ln \ubar{\th}$. If the valuation $\th$ is slightly below $\ubar{\th}$, however, the buying probability drops to $0$ and the regret, $- (u \circ \hat{f})(\th)$, jumps up from $R$ to $\th$.

\subsection{Bayesian persuasion} \label{sec:BP}

\begin{figure}
\centering
    \begin{tikzpicture}
    \begin{axis}[ 
        xmin = 0, xmax = 1.1,
        ymin = 0, ymax = 1.1,
        axis lines = middle,
        xtick = {0.3, 0.4, 0.6,1},
        xticklabels = {$\a$, $\mu$, $\b$, $1$},
        ytick = {0.6, 1}, 
        yticklabels = {$2 \a$, 1},
        clip = false, 
        xlabel = {$\th$},
	xlabel style={at=(current axis.right of origin), anchor=west},
    ]

    \addplot[very thick, Blue]  coordinates {(0.3,0.6) (1,1)} node[pos = 0.8, above, xshift = -3pt, yshift = 4pt] {$u \circ \hat{f}_\a$};
    \addplot [very thick, Blue] coordinates {(0,0) (0.3,0)};
    \addplot [dashed, Blue] (0.3,0) -- (0.3,0.6);

    \filldraw [Blue] (0.3,0.6) circle (2pt);
    \filldraw [Blue, fill =white] (0.3,0) circle (2pt);

    \draw (0.3,0) node[cross=3pt, thick, Orange]{};
    \draw (0.6,0) node[cross=3pt, thick, Orange]{};
\end{axis}
\end{tikzpicture}
\caption{Bayesian persuasion with fixed prior mean and support restriction}
\label{fig:KG}
\end{figure}

\cite{HuWeng2021} consider a maxmin version of the persuasion problem in \cite{KamenicaGentzkow2011}.\footnote{We focus on a special case of \citeapos{HuWeng2021} results. \cite{Kosterina2022} studies a maxmin version of a continuous persuasion problem. \cite{DworczakPavan2022} consider a different robustness notion for persuasion problems. In their model, Nature can give the receiver additional information, contingent on the realization of the sender's experiment.} The sender commits to a Blackwell experiment about a binary fundamental $\w \in \W = \{0,1\}$. The receiver observes the realization of the experiment and chooses a binary action $a \in \{0,1\}$. Payoffs for the sender and receiver are given by $u_S(a, \w) = a$ and $u_R(a,\w) = -(a - \w)^2$. The sender is uncertain of the receiver's belief over $\W$. Thus, the state $\th \in \Th = [0,1]$ is the receiver's belief, i.e., the probability assigned to $\w = 1$. The ambiguity set $\Pi$ contains all priors with fixed mean $\mu$ that concentrate on the interval $[\a, \b]$, where $0 < \a < \mu < \b \leq 1$ and $\a < 1/2$. The interpretation is that the sender and receiver initially have common belief $\mu$, but the sender is uncertain of what additional information the receiver gets. The sender knows only that the receiver's belief remains in the interval $[\a, \b]$. 

\citet{HuWeng2021} show that if $\mu$ is sufficiently close to $\a$, then it is maxmin-optimal to use the binary experiment from \cite{KamenicaGentzkow2011} that splits the belief $\a$ between $0$ and $1/2$.\footnote{See Proposition  3 (p.~928) and Proposition 4 (p.~930). This experiment is uniquely optimal if $\a < \b \leq 1/2$ or $\a < 1 - \b < 1/2 < \b$.} Let $\hat{f}_\a$ denote the social choice function induced by this experiment. The value function $u \circ \hat{f}_\a$ is shown in \cref{fig:KG}. The probability of the ``high'' signal realization is $\th + (1 - \th) \a / (1 - \a)$. This realization induces the receiver to choose action $a = 1$ if and only if $\th \geq \a$.  The payoff guarantee from $\hat{f}_\a$ over $\Pi$ is not robust. Consider the prior $\pi = p \d_{\a} + (1 -p) \d_{\b}$, where $p \a + (1 - p) \b = \mu$. The point masses on $\a$ and $\b$ are indicated on the plot. The prior $\pi$ is in $\Pi$. But if $\pi$ is perturbed to $\pi_\e = p \d_{\a - \e} + (1 -p) \d_{\b}$, for any $\e$ in $(0,\a)$, then the sender's expected utility from $\hat{f}_\a$ drops by $2 \a p$.

\subsection{Multi-good monopoly problem} \label{sec:Carroll}

\cite{Carroll2017} considers a maxmin version of the multi-good monopoly problem. There are $K$ goods, where $K \geq 2$. Fix a valuation upper bound $M > 0$. The agent's type $\th = (\th_1, \ldots, \th_K) \in \Th = [0,M]^K$ specifies his valuation for each of the $K$ goods. The agent's valuation for a bundle of goods equals the sum of the valuations of the goods in the bundle. The principal knows the marginal distribution of the agent's valuation for each good, but not the joint distribution of the valuations. Formally, the principal seeks a revenue guarantee over all joint distributions with fixed marginals $\pi_1, \ldots, \pi_K$. By \citet[Theorem 2.1, p.~458]{Carroll2017}, it is maxmin-optimal for the principal to screen the agent separately along each dimension, i.e., to post for each good $k$ a price $\hat{p}_k \in \argmax_{p} p \cdot \pi_k ([ p, M])$.  The value function induced by this mechanism is continuous outside of the set $\cup_{k=1}^{K} \{ \th \in \Th: \th_k = \hat{p}_k \}$. If $\pi_k ( \{ \hat{p}_k\} ) = 0$ for each $k$, then the payoff guarantee from this mechanism is robust (by \cref{res:robustness_continuity}.\ref{it:closure}).  On the other hand, if there is some $k$ such that $\pi_k ( \{ \hat{p}_k\} ) > 0$ and $\hat{p}_k > 0$, then the payoff guarantee is not robust; the reasoning is similar to the monopoly selling example with a median constraint (\cref{sec:monopoly}).

\subsection{Delegated project choice} \label{sec:delegated_project_choice}

\begin{figure}
\centering
\begin{tikzpicture}
    \begin{axis}[
        xmin = 0, xmax = 1.1,
        ymin = 0, ymax = 1.1,
        axis lines = middle,
        xtick = {0.25,1},
        xticklabels = {$\ubar{u}_A$, $1$},
        ytick = {4/7,1},
        yticklabels = {$1-R$,$1$},
        clip = false, 
        xlabel = {$u_A$},
	xlabel style={at=(current axis.right of origin), anchor=west},
        ylabel = {$u_P$},
	ylabel style={at=(current axis.north west), anchor=south}, 
    ]

    \coordinate (A) at (0.25,0);
    \coordinate (B) at (0.25,1);
    \coordinate (C) at (1,1);
    \coordinate (D) at (1,0);
    \coordinate (E) at (1,4/7);
    \coordinate (F) at (0.25,4/7);

 
    \draw[fill=Blue!25] (A) -- (B) -- (C) -- (D) -- cycle;

    \draw[fill=Blue!50] (B) -- (C) -- (E) -- (F) -- cycle;

    \addplot[thick] coordinates { (0.25,4/7) (1,4/7)};

    \draw (0.25,1) node[cross=4pt, very thick, Orange]{};
    \draw (1,0) node[cross=4pt, very thick, Orange]{};

\end{axis}
\end{tikzpicture}
\caption{Delegated project choice: two-tiered mechanism (blue) and a feasible set (orange)}
\label{fig:GS}
\end{figure}

\cite{GuoShmaya2023} consider a maxmin version of the delegated project choice problem in \cite{ArmstrongVickers2010}. There are two players: a principal (she) and an agent (he). A project is a pair $u = (u_A,u_P) \in \mathcal{U}$ indicating the payoffs of the agent and the principal; here, $\mathcal{U}$ is a compact subset of $\R^2$ that includes a neighborhood of $[0,1]^2$. Payoffs are normalized relative to the status quo $(0,0)$. The agent privately knows the nonempty, compact set $\mathcal{A} \subseteq \UU$ of available projects. The principal must select an available project or the status quo. This problem fits into our framework by taking the state to be the available project set $\AA$.\footnote{\label{ft:subset_Polish}Thus, the state space contains all nonempty, compact subsets of $\UU$. We endow this space with the Hausdorff metric topology (with respect to the usual Euclidean topology on projects). The space of compact subsets of a Polish space inherits a unique Hausdorff metric topology \citep[3.91, p.~120]{AliprantisBorder2006}, and this topology is itself Polish by \citet[Theorem 2.5.3, p.~72]{Edgar2008} and \citet[p.~120]{AliprantisBorder2006}.} The principal's regret from project $u$ in state $\AA$ is $\max_{u' \in \AA \cup \{ (0,0)\}} u_P' - u_P$; this is the principal's forgone utility relative to her favorite project in $\AA \cup \{ (0,0) \}$.

The agent can propose one project $u$ from  $\mathcal{A}$.\footnote{It is assumed that it is infeasible for the agent to propose a project outside $\AA$. The single-proposal protocol is the simplest case considered in \cite{GuoShmaya2023}. They also solve the case of multiple proposals.} The principal commits to a mechanism $\a \colon \UU \to [0,1]$. Under mechanism $\a$, the principal adopts any proposal $u$ with probability $\a (u)$. With complementary probability, the principal keeps the status quo $(0,0)$. A mechanism and an associated best response from the agent induce a social choice function, which specifies in each state $\AA$ a lottery over projects. The principal can select among the agent's best responses in the case of a tie. The principal evaluates each social choice function according to its worst-case (expected) regret over all finite project sets $\AA \subseteq [\ubar{u}_A, 1] \times [0, 1]$, where the parameter $\ubar{u}_A$ is in $[0,1]$. By \citet[Theorem 4.1, p.~1576]{GuoShmaya2023}, the optimal regret guarantee is $R = (1 - \ubar{u}_A)/(2 - \ubar{u}_A)$, and the following mechanism is maxmin optimal. \cref{fig:GS} illustrates the mechanism. A project $u$ is \emph{top-tier} if $u_P \geq 1- R$. If the agent proposes a top-tier project (in the dark shaded region), then the principal adopts the proposal with certainty. Any other proposal $u$ (in the light shaded region) is adopted with probability $\ubar{u}_A / u_A$ (where $0/0 = 1$). If the agent proposes a project that is not top-tier, then his expected utility is exactly $\ubar{u}_A$. Thus, the agent finds it optimal to propose some top-tier project if any top-tier project is available.

If $\ubar{u}_A > 0$, then the regret guarantee from the proposed mechanism is not robust.\footnote{To fully specify this mechanism, the function $\a$ must be extended from $[\ubar{u}_A,1] \times [0,1]$ to $\UU$, but our argument goes through for any such extension.} Consider the set $\AA = \{( \ubar{u}_A, 1), (1,0)\}$ marked on the graph in \cref{fig:GS}. In state $\AA$, it is optimal for the agent to propose the top-tier project $(\ubar{u}_A, 1)$. This project is adopted, so the principal's regret is $0$. If, however, the set $\AA$ is perturbed to $\AA_\e = \{ (\ubar{u}_A - \e, 1), (1,0) \}$, for any $\e \in (0, \ubar{u}_A)$, then the agent strictly prefers to propose the project $(1,0)$. This project is adopted with probability $\ubar{u}_A$, and the status quo $(0,0)$ is adopted with probability $1 - \ubar{u}_A$. Thus, the principal's regret equals $1$, which is strictly worse than the regret guarantee $R$. In fact, a regret of $1$ is weakly worse than the regret from \emph{any} mechanism in any state $\AA \subseteq [\ubar{u}_A, 1] \times [0, 1]$.

\subsection{Moral hazard}

In an influential paper, \cite{Carroll2015} considers a robust version of the classic moral hazard problem. The agent privately chooses an action, which generates a stochastic output. The principal observes the realized output, but not the agent's action. There is a compact set $\YY \subseteq \R_+$ of possible output levels, with $\min \YY = 0$. An action for the agent is represented by a pair $(F,c) \in \D(\YY) \times \R$, where $F$ is the  output distribution and $c$ is the agent's cost. There is a nonempty, compact set $\AA$ of actions that are available. This set $\AA$, called the \emph{technology}, is privately known to the agent. To represent this problem in our framework, we take the state to be the technology $\AA$.\footnote{The space $\D(\YY)$ is endowed with the weak topology. The space $\D(\YY) \times \R$ is endowed with the associated product topology, which is Polish. The space of nonempty, compact subsets of $\D(\YY) \times \R$ is endowed with the Hausdorff metric topology, which is itself Polish; see \cref{ft:subset_Polish}.} 

The principal chooses a contract, which is a continuous function $w \colon \YY \to \R_+$ specifying the agent's nonnegative wage as a function of the realized output. Given a contract $w$, the agent chooses an action from the technology $\AA$ to maximize his expected utility. A contract and an associated best response from the agent induce a social choice function, which specifies in each state $\AA$ a lottery over the output and the agent's wage and cost. There is a nonempty, compact set $\AA_0$ of actions that the principal knows are available. The principal does not know which other actions are available (but she does know that all actions have nonnegative cost). The principal evaluates each social choice function according to its worst-case (expected) profit over all technologies $\AA$ satisfying $\AA_0 \subseteq \AA \subseteq \D( \YY) \times \R_+$. To rule out the trivial case, it is assumed that there exists $(F,c)$ in $\AA_0$ such that $\E_F[y]  > c$.

\citet[Theorem 1, p.~541]{Carroll2015} shows that there exists a linear contract that is maxmin optimal.\footnote{The class of contracts rules out screening on the technology (by offering the agent a menu of contracts to choose from) and randomization (where the agent is informed of the contract realization before choosing his action). \citet[Theorem 4, p.~550]{Carroll2015} shows that screening cannot increase the principal's payoff guarantee. On the other hand, \cite{KTV2025} show that randomization can strictly improve the principal's payoff guarantee. Furthermore, they show that in the class of all random contracts, it is optimal to randomize over a continuum of \emph{linear} contracts.} The intuition is as follows \citep[p.~537]{Carroll2015}:
\begin{quote}
When the principal proposes a contract, in the face of her uncertainty about the agent’s technology, she knows very little about what will happen. But the one thing she does know is a lower bound on the agent’s expected payoff (from the actions that are known to be available). The only effective way to turn this into a lower bound on her own expected payoff is via a linear relationship between the two.
\end{quote}
The \emph{agent}'s expected payoff from any contract is continuous in the technology. For any maxmin-optimal linear contract that is nonzero, the principal's payoff guarantee is robust because this guarantee is derived from a linear, hence continuous, function of the agent's payoff; see \cref{res:robust_guarantee} in \cref{sec:linear_contracts}. Note, however, that the \emph{principal's} expected payoff from a linear contract is not continuous in the technology---the principal's payoff can jump when the agent's preferred action changes. Therefore, \cref{res:robust_guarantee} cannot be proven by directly applying \cref{res:robustness_continuity}. Indeed, if the zero contract is maxmin-optimal, then its payoff guarantee is not robust, as we show in \cref{sec:linear_contracts}.

\cite{DurandardVaidyaXu2025wp} consider a different robust contracting problem in which the agent engages in sequential search. 
Contracts and technologies are defined as in \cite{Carroll2015}, but now each pair $(F,c)$ is interpreted as a ``box'' in the sense of \cite{Weitzman1979}; the agent can pay cost $c$ to open the box and observe the realization from $F$. Given a collection $\AA$ of boxes, the agent sequentially chooses which boxes to open. Then the agent chooses which of the observed output realizations to submit to the contract $w$. There is one box $(F_0, c_0)$ that the principal knows is available, but the principal does not know which other boxes are available. To rule out the trivial case, it is assumed that $\E_{F_0} [y] - c_0 > 0$. 

\cite{DurandardVaidyaXu2025wp} show that the principal can guarantee a payoff of $\E_{F_0} [y] -c_0$, thus fully extracting the surplus from the known box.\footnote{This payoff guarantee is higher than the corresponding guarantee in \cite{Carroll2015} in the case $\AA_0 = \{ (F_0, c_0) \}$. The lower-cost, lower-output actions that reduce the payoff guarantee in \cite{Carroll2015} do not have the same effect in the sequential search problem because the agent can open multiple boxes, thus increasing output.} Each maxmin-optimal contract takes a debt-like form in which the agent's wage is zero whenever output is below a threshold. But the payoff guarantee from each maxmin-optimal contract fails to be robust. For any $\e > 0$, if there is a single box $(F_0, c_0 + \e)$, then the agent strictly prefers not to open it, and hence the principal's expected profit drops to $0$.

\section{Non-robustness and finite-support worst-case priors} \label{sec:saddle} 

To explain the failure of robustness in some of the examples from \cref{sec:fragile}, we illustrate how saddle points with finite-support priors typically yield ``simple'' maxmin-optimal mechanisms with non-robust payoff guarantees.  

The standard approach to deriving a maxmin-optimal mechanism is to solve for a saddle point of the zero-sum game between the designer and Nature. In this game, the designer chooses a social choice function from the feasible set $\FF$, and Nature simultaneously chooses a prior from the ambiguity set $\Pi$. Formally, a pair $(\hat{f}, \hat{\pi}) \in \mathcal{F} \times \Pi$ is a \emph{saddle point} if for all $f \in \FF$ and $\pi \in \Pi$, we have
\begin{equation} \label{eq:saddle_def}
  \ang{u \circ f, \hat{\pi}} \leq  \ang{ u \circ \hat{f}, \hat{\pi}} \leq \ang{u \circ \hat{f}, \pi}.
\end{equation}
The first inequality says that $\hat{f}$ is Bayesian optimal under the prior $\hat{\pi}$. The second inequality says that the prior $\hat{\pi}$ minimizes over $\Pi$ the designer's expected payoff from $\hat{f}$. If $(\hat{f}, \hat{\pi})$ is a saddle point, then $\hat{f}$ is maxmin optimal, and the payoff guarantee from $\hat{f}$ over $\Pi$ equals $\ang{ u \circ \hat{f}, \hat{\pi}}$, the value of the zero-sum game.

Consider a fixed prior and an associated Bayesian-optimal social choice function. We first characterize the range of ambiguity sets for which this pair constitutes a saddle point. We need a few definitions. For any social choice function $\hat{f}$ and prior $\hat{\pi}$, let
\[
    \Pi_0 ( \hat{f}, \hat{\pi}) = \Set{ \pi \in \D(\Th) : \ang{u \circ \hat{f}, \pi} \geq \ang{u \circ \hat{f}, \hat{\pi}} }.
\]
The set $\Pi_0 ( \hat{f}, \hat{\pi})$ is the intersection of $\D(\Th)$ with an (algebraic) half-space. Note that $\Pi_0 ( \hat{f}, \hat{\pi})$ contains the prior $\hat{\pi}$. To emphasize that the definition of a saddle point depends on the ambiguity set $\Pi$, we refer to a $\Pi$-saddle point below. 

\begin{prop}[Bayesian solutions as saddle points] \label{res:Bayesian_robustness} Fix a feasible set $\FF$. Let $\hat{\pi}$ be a prior in $\D(\Th)$. Let  $\hat{f}$ be a social choice function that is Bayesian optimal under $\hat{\pi}$ (with respect to $\FF$). For any nonempty subset $\Pi$ of $\D(\Th)$, the following are equivalent: 
\begin{enumerate}[label = (\roman*), ref = \roman*]
    \item $(\hat{f}, \hat{\pi})$ is a $\Pi$-saddle point; 
    \item $\{ \hat{ \pi} \} \subseteq \Pi \subseteq \Pi_0 ( \hat{f}, \hat{\pi})$.
\end{enumerate}
Moreover, $\Pi_0 ( \hat{f}, \hat{\pi}) \neq \{ \hat{\pi} \}$ unless there exists $\th \in \Th$ such that $\hat{\pi} = \d_{\th}$ and $\{\th \} = \argmax_{\th' \in \Th} (u \circ \hat{f}) (\th')$.
\end{prop}

Mathematically, \cref{res:Bayesian_robustness} is immediate from the definition of a saddle point. Conceptually, it reveals the range of ambiguity sets that will deliver a given Bayesian-optimal mechanism as a maxmin solution. Indeed, every mechanism that is Bayesian-optimal with respect to some prior is also maxmin optimal with respect to some ambiguity set that is  a half-space. 

We illustrate the construction from \cref{res:Bayesian_robustness} in the monopoly selling problem. For any integrable prior $\hat{\pi}$, there exists a posted price $\hat{p}$ that is Bayesian optimal. Let $\hat{f}$ denote the social choice function induced by this posted price $\hat{p}$. We have
\[
    \Pi_0 ( \hat{f}, \hat{\pi}) = \Set{ \pi  \in \D(\Th) : \pi ( [ \hat{p}, \infty)) \geq \hat{\pi}([\hat{p}, \infty)) }.
\]
Let $\a = 1 - \hat{\pi}  ( [\hat{p}, \infty))$. Note that $\Pi_0 ( \hat{f}, \hat{\pi})$ is the set of valuation distributions with (largest) $\a$-quantile at least $\hat{p}$. If $\a = 1/2$, then $\Pi_0 ( \hat{f}, \hat{\pi})$ is defined by a restriction on the median, similar to the ambiguity set in the example from the introduction.

Building on \cref{res:Bayesian_robustness}, we give a recipe for mechanically obtaining a maxmin-optimal mechanism that is simple (in the sense that it can be implemented with a small menu). Consider a single-agent screening problem, where $\Th$ is the set of agent types, and the feasible set $\FF$ contains all social choice functions satisfying incentive compatibility and participation. Fix a prior $\hat{\pi} \in \D(\Th)$ that is supported on $m$ types. Suppose that under the prior $\hat{\pi}$, the Bayesian design problem has a solution. Then, by the taxation principle, there exists a Bayesian-optimal social choice function that can be implemented with a menu of at most $m$ alternatives in addition to the outside option. By \cref{res:Bayesian_robustness}, there is a range of ambiguity sets $\Pi$ with respect to which $\hat{f}$ is maxmin optimal. But the simplicity of $\hat{f}$ is not due to the maxmin objective per se. Rather, $\hat{f}$ is simple because the Bayesian solution takes a simple form at the particular prior $\hat{\pi}$. 

We now illustrate why the payoff guarantee from this construction is not typically robust. We formally establish this non-robustness in a \emph{profit-maximization problem}, which is defined as follows. The state space $\Th$ is the set of agent types. Let $Q$ be a measurable allocation space. The principal chooses an allocation $q \in Q$ and a transfer $t \in \R$. Thus, the decision space is $X = Q \times \R$. The agent's utility is $v( q, \th) - t$, where $v \colon Q \times \Th \to \R_+$ is bounded and measurable. The principal's utility is the profit $u(q,t,\th) = t - c(q)$, where the cost function $c \colon Q \to \R_+$ is bounded and measurable. The feasible set $\FF$ contains all social choice functions $f \colon \Th \to \D (X)$ that satisfy incentive compatibility and participation constraints, where the agent's outside option utility is normalized to zero. To represent the outside option, we assume that there exists an allocation $q_0 \in Q$ such that $c (q_0) = 0$ and $v(q_0, \th) = 0$ for all $\th$ in $\Th$. Technically, the principal's utility function is not bounded, but the value function $u \circ f$ is bounded for each social choice function $f$ in $\FF$ that is not strictly dominated (for the principal); this is shown in the proof of \cref{res:saddle} in \cref{sec:saddle_proof}.

As part of the definition of the profit-maximization problem, we impose a fullness assumption on the type space.  For any types $\th, \th' \in \Th$, type $\th'$ is \emph{weaker} than type $\th$ if for each allocation $q \in Q$, we have $v(q, \th') \leq v(q,\th)$, with strict inequality whenever $v(q,\th) > 0$. In particular, type $\th$ is weaker than itself if and only if $v(q,\th) = 0$ for all $q \in Q$.  We assume that for each type $\th$, every neighborhood of $\th$ contains some type weaker than $\th$. This assumption is satisfied in classical single-dimensional screening settings and in the multi-good monopoly problem (with additive valuations), provided that the space of valuations is not bounded away from zero. 

\begin{prop}[Saddle points with finite-support priors are not robust] \label{res:saddle}
Fix a profit-maximization problem, $(\Th, X, u, \FF)$, and an ambiguity set $\Pi$. Let $\hat{\pi}$ be a finite-support prior in $\Pi$ and let $\hat{f}$ be a social choice function in $\FF$. If $(\hat{f}, \hat{\pi})$ is a saddle point and $\ang{ u \circ \hat{f}, \hat{\pi}} > 0$, then the payoff guarantee from $\hat{f}$ over $\Pi$ is not robust.  
\end{prop}

To prove \cref{res:saddle}, we generalize the logic from some of the examples in \cref{sec:fragile}. Consider a saddle point $(\hat{f}, \hat{\pi})$ with $\ang{ u \circ \hat{f}, \hat{\pi}} > 0$. First, we show that there exists some type $\th_0$ in the support of $\hat{\pi}$ such that $(u \circ \hat{f}) (\th_0) > 0$ and the participation constraint for type $\th_0$ holds with equality. Otherwise, the principal could strictly increase profits, contrary to the optimality of $\hat{f}$. Next, we show that under $\hat{f}$, the principal does not earn strictly positive profit from any type weaker than $\th_0$. Therefore, if probability $\hat{\pi}(\th_0)$ is moved from $\th_0$ to any arbitrarily close weaker type, the designer's expected profit from $\hat{f}$ drops by at least $\hat{\pi} (\th_0) (u \circ \hat{f}) (\th_0)$. Thus, the payoff guarantee from $\hat{f}$ over $\Pi$ is not robust. 

\section{Obtaining robust payoff guarantees} \label{sec:classification}

We turn to positive results about robustness. First, we show that robustness is assured whenever the ambiguity set is \emph{rich}, a property satisfied by some commonly used ambiguity sets. For ambiguity sets that are not rich, we introduce an algorithm for constructing a maxmin-optimal mechanism with a robust payoff guarantee. We illustrate our algorithm in the regret-minimizing monopoly selling problem from \cite{BergemannSchlag2008}.

\subsection{Uniform robustness}

We define a strong property of an ambiguity set---uniform robustness---which implies that every mechanism's payoff guarantee over that set is robust. To define this property, it is helpful to decompose our setting $(\Th, X, u)$ into the state space $\Th$ and the \emph{decision environment} $(X,u)$. Recall that in any decision environment, the utility function $u$ is assumed to be bounded. 

\begin{defn}[Uniform robustness] Fix the state space $\Th$. A nonempty subset $\Pi$ of $\D(\Th)$ is \emph{uniformly robust} if in every decision environment $(X,u)$ the following holds: for every sequence $(\pi_n)$ in $\D(\Th)$ that converges to a prior in the closure of $\Pi$, 
\begin{equation} \label{eq:uniform}
 \liminf_n \Paren{ \inf_{f} \Brac{ \ang{u \circ f, \pi_n} - \inf_{\pi \in \Pi} \ang{u \circ f, \pi} }}\geq 0,
\end{equation}
where the infimum is over all social choice functions $f \colon \Th \to \D(X)$.
\end{defn}

Recall that robustness was defined as a property of a pair $(f, \Pi)$. Uniform robustness, by contrast, is a property of the ambiguity set $\Pi$ alone. Intuitively, an ambiguity set is uniformly robust if in every decision environment, the violation of the payoff guarantee at priors sufficiently near the ambiguity set is controlled uniformly over all social choice functions.  If the designer uses a uniformly robust ambiguity set, then she is assured that whichever decision environment she faces and whichever social choice function she implements, the associated payoff guarantee will be robust. In particular, any maxmin-optimal mechanism's payoff guarantee will be robust. Uniform robustness is useful even in applications where the utility function $u$ is not bounded: If $\Pi$ is uniformly robust, then for any social choice function $f$ for which $u \circ f$ is bounded, the payoff guarantee from $f$ over $\Pi$ is robust.

We identify a structural property of an ambiguity set that implies uniform robustness. Let $\| \cdot \|_{\mathrm{TV}}$ denote the total variation norm on $\D(\Th)$.\footnote{Given $\pi, \pi' \in \D(\Th)$, let $\| \pi - \pi' \|_{\mathrm{TV}}  = \sup_A | \pi (A) - \pi'(A)|$, where the supremum is over all Borel subsets $A$ of $\Th$.} 

\begin{defn}[Richness] An ambiguity set $\Pi$ is \emph{rich} if for every sequence $(\pi_n)$ in $\D(\Th)$ that converges to a prior in the closure of $\Pi$, there exists a sequence $(\pi_n')$ in $\Pi$ such that $ \| \pi_n' - \pi_n \|_{\mathrm{TV}} \to 0$.
\end{defn}

Richness requires, roughly, that whenever a sequence of priors approaches the set $\Pi$ in the weak topology, this sequence also approaches $\Pi$ in the total variation norm. To illustrate this definition, we contrast two ambiguity sets in the monopoly selling setting with $\Th = \R_+$. Fix $\l > 0$. Let $\Pi_1$ contain all priors with \emph{mean} $\l/2$. Let $\Pi_2$ contain all priors that have $\l$ as a \emph{median}. Both sets contain the prior $\pi = \d_0/2 + \d_\l / 2$. Let $(\l_n)$ be a strictly increasing sequence that converges to $\l$. For each $n$, let $\pi_n = \d_0/2 + \d_{\l_n}/2$. The sequence $(\pi_n)$ weakly converges to $\pi$, but $\| \pi_n - \pi\|_{\mathrm{TV}} = 1/2$ for all $n$. For each $n$, let
\[
\e_n = \frac{ \l - \l_n}{1 + \l - \l_n}
\quad
\text{and}
\quad
\pi_n' = \pi_n + \frac{\e_n}{2} (\d_{\l + 1} - \d_{\l_n}).
\]
In words, $\pi_n'$ is obtained from $\pi_n$ by moving probability $\e_n/2$ from state $\l_n$ to state $\l + 1$. The resulting prior $\pi_n'$ has mean $\l/2$ and hence is in $\Pi_1$. The probability $\e_n/2$ converges to $0$ as $n \to \infty$. On the other hand, no such reallocation of a small amount of probability can produce a prior in $\Pi_2$. Suppose that some probability strictly less than $1/2$ is moved from $\pi_n$. After this modification, the probability assigned to the set $[0, \l_n]$ is still strictly greater than $1/2$, so the median is strictly below $\l$. Thus, the modified distribution is not in $\Pi_2$. 

\begin{prop}[Richness implies uniform robustness] \label{res:rich} Every rich ambiguity set is uniformly robust. 
\end{prop}

Here is a sketch of the proof. Consider a rich ambiguity set $\Pi \subseteq \D(\Th)$, a decision environment $(X,u)$, and a social choice function $f$. Let $(\pi_n)$ be a sequence of priors that converges to some prior $\pi$ in the closure of $\Pi$. The associated payoffs $\ang{ u \circ f, \pi_n}$ do not necessarily converge to $\ang{ u \circ f, \pi}$ because  $u \circ f$ may not be continuous. In particular, the payoffs $\ang{ u \circ f, \pi_n}$ may be uniformly lower than $\ang{ u \circ f, \pi}$. But because $\Pi$ is rich, there is another sequence $(\pi_n')$ of priors in $\Pi$ such that $ \| \pi_n' - \pi_n \|_{\mathrm{TV}} \to 0$. By the properties of the total variation norm, the difference between the payoffs $\ang{ u \circ f, \pi_n'}$ and $\ang{u \circ f, \pi_n}$ is bounded by a constant multiple of $\| \pi_n' - \pi_{n}\|_{\mathrm{TV}}$ that is independent of $f$. Since $\inf_{\pi \in \Pi} \ang{u \circ f, \pi} \leq \ang{ u \circ f, \pi_n'}$ for each $n$, condition \eqref{eq:uniform} follows.

\subsection{Common ambiguity sets that are uniformly robust} \label{sec:common_ambiguity}

We formally define two commonly used classes of ambiguity sets. Then we show that these sets are rich, and hence, by \cref{res:rich}, 
uniformly robust.

\paragraph{Moment sets} Recall the moment restrictions in \citeapos{CarrascoEtal2018} maxmin specification of the monopoly selling problem, described in \cref{sec:monopoly}.\footnote{Moment restrictions have also been used to analyze the newsvendor problem \citep{Scarf1958} and an auction problem \citep{AzarMicali2012}.} Here, we define moment restrictions in our general setting. For any measurable function $g \colon \Th \to \R^m$ and any subset $Y$ of $\R^m$, let
\[
    M(g,Y) = \Set{ \pi \in \D( \Th): \E_{\th \sim \pi} [g(\th)] \in Y},
\]
where the constraint $\E_{\th \sim \pi} [g(\th)] \in Y$ implicitly requires that $|g_j|$ is $\pi$-integrable for each component $j$.

We are interested in restrictions on \emph{continuous} moment functions. Even if $g$ is continuous, there are pathological examples in which $M(g,Y)$ encodes a restriction on a \emph{discontinuous} moment function. For example, let $\Th = \R_+$ and fix $\th_0 > 0$. Let $g(\th) = (\th -\th_0)^2$ and $Y = ( -\infty, 0]$. Then $M (g, Y)$ equals $\{ \d_{\th_0} \}$, which should not count as a continuous moment set. The problem here is that $Y$ does not contain any interior point of the image $g(\Th) = [ 0, \infty)$. We impose conditions on $Y$ to rule out such examples. 

For any subset $S$ of $\R^m$, let $\conv S$ denote the convex hull of $S$. A subset $Y$ of $\R^m$ is \emph{star-$g$-interior} if there exists a point $y_0$ in the relative interior of $\conv g (\Th)$ such that $[y_0, y] \subseteq  Y \cap \conv g(\Th)$ for all $y \in Y \cap \conv g(\Th)$.  An ambiguity set $\Pi$ is a \emph{continuous moment set} if $\Pi = M(g, Y)$ for some dimension $m \geq 1$, some continuous function $g \colon \Th \to \R^m$, and some star-$g$-interior subset $Y$ of $\R^m$. It can be verified that the ambiguity set from \cite{CarrascoEtal2018} is a continuous moment set in this sense.

\paragraph{Metric neighborhoods} \label{page:metric} Another natural choice of ambiguity set is a neighborhood of some reference prior or set of priors.\footnote{In the monopoly selling problem, \cite{BergemannSchlag2011} consider a Prokhorov ball. In a contracting problem, \cite{Auster2018} uses an $\e$-contamination neighborhood.} Here, we define metric neighborhoods in our general setting. Given a metric $D$ on $\D(\Th)$, a radius $r > 0$, and a nonempty subset $C$ of $\D(\Th)$, the  \emph{metric-$D$, radius-$r$ neighborhood of $C$} is defined by
\[
    N_D( C,r)  = \Set{ \pi \in \D( \Th): D(\pi, C) \leq r},
\]
where $D(\pi, C) = \inf_{\pi' \in C} D ( \pi, \pi')$. An ambiguity set $\Pi$ is a \emph{neighborhood with respect to $D$} if $\Pi = N_D ( C, r)$ for some radius $r > 0$ and some nonempty subset $C$ of $\D(\Th)$.

A metric on $\D(\Th)$ is \emph{weak} if it metrizes the weak topology on $\D(\Th)$. A metric on $\D(\Th)$ is \emph{convex} if it is a convex function on its domain $\D(\Th) \times \D(\Th)$. In particular, any metric induced by a norm is convex. A leading example of a weak, convex metric is the Wasserstein metric, which we now define.\footnote{Other examples of weak, convex metrics include those induced by the Kantorovich--Rubinstein and Fortet--Mourier norms \cite[p.~109]{Bogachev2018}. On the other hand, the Prokhorov metric is not convex. Indeed, there exist Prokhorov balls that are not uniformly robust.} A metric on the Polish space $\Th$ is \emph{compatible} if it metrizes the given topology on $\Th$. Given a bounded, compatible metric $d$ on $\Th$, the \emph{Wasserstein metric} $W$ on $\D(\Th)$ is defined by
\begin{equation} \label{eq:W}
    W(\mu, \nu) = \inf_{\g} \E_{(\th, \th') \sim \g} \Brac{ d(\th, \th') },
\end{equation}
where the infimum is over all probability measures $\g$ in $\D ( \Th \times \Th)$ with first marginal $\mu$ and second marginal $\nu$. Thus, $W( \mu, \nu)$ is the infimal expected moving distance when transporting mass from $\mu$ to $\nu$. 

\begin{prop}[Rich ambiguity sets] \label{res:robust_sets} 
All ambiguity sets taking the following forms are rich:
\begin{enumerate}[label = (\roman*), ref = \roman*]
    \item continuous moment sets;
    \item neighborhoods with respect to any weak, convex metric.
\end{enumerate}
\end{prop}

\begin{rem}[Wasserstein distance with an unbounded metric] Given an \emph{unbounded}, compatible metric $d$ on $\Th$, we can still define the function $W$ by the formula in \eqref{eq:W}. In this case, $W$ can take the value $\infty$, so $W$ is technically an extended metric. Nevertheless, neighborhoods with respect to $W$ are well-defined, and these neighborhoods are rich, as we check in the proof of \cref{res:robust_sets}.
\end{rem}

Here is a sketch of the proof of \cref{res:robust_sets}. Consider an ambiguity set $\Pi$ taking one of these two forms. For any sequence $(\pi_n)$ of priors converging to some prior in the closure of $\Pi$, we construct a sequence $(\pi_n')$ in $\Pi$ such that $\| \pi_n' - \pi_n \|_{\mathrm{TV}} \to 0$. For each $n$, we define $\pi_n'$ to be a mixture of $\pi_n$ (or a total-variation norm approximation of $\pi_n$) with some carefully chosen prior $\z_n$. If $\Pi$ is a metric neighborhood $N_D(C,r)$, then we choose $\z_n$ in $C$. If $\Pi$ is a continuous moment set $M(g,Y)$, then we choose $\z_n$ so that the resulting mixture has the correct moments. The weight on $\z_n$ in $\pi_n'$ tends to $0$ as $n \to \infty$. 

\subsection{Recovering robustness} \label{sec:algorithm}

After solving for a maxmin-optimal mechanism, we propose that the analyst check whether the associated payoff guarantee is robust. If the guarantee is not robust, then there are various ways to proceed. We first present an algorithm that recovers robustness by enriching the ambiguity set in the original maxmin objective. Then we relate our algorithm to other approaches that move beyond maxmin preferences.

Given the original ambiguity set $\Pi$, the algorithm proceeds as follows. Recall the properties of metrics defined in \cref{sec:common_ambiguity}. Choose a weak, convex metric $D$ on $\D(\Th)$ and a radius $r > 0$. Let $\Pi'$ be the metric neighborhood $N_D ( \Pi, r)$. Compute a maxmin-optimal mechanism with respect to the new ambiguity set $\Pi'$.\footnote{It may be tempting to keep the original ambiguity set $\Pi$ but restrict the maximization to social choice functions whose payoff guarantees over $\Pi$ are robust. This procedure has serious shortcomings, however. Even if the restricted collection of social choice functions is nonempty (which is not guaranteed), one can construct examples in which the maximizing social choice function is  strictly dominated by a social choice function whose payoff guarantee over $\Pi$ is not robust.}  The ambiguity set $\Pi'$ is uniformly robust (by \cref{res:rich,res:robust_sets}), so the new mechanism's payoff guarantee over $\Pi'$ is robust. Moreover,  every social choice function's payoff guarantee over $N_D (\Pi, r)$ is continuous as a function of the radius $r \in (0, \infty)$; see \cref{res:continuity_radius}  in \cref{sec:continuity_in_radius}. This property distinguishes our approach from other ways of enlarging the ambiguity set, as we illustrate in an example below. 

As long as the original ambiguity set $\Pi$ is convex,\footnote{If $\Pi$ is not convex, one can first replace $\Pi$ with its convex hull, $\conv \Pi$. Then the new ambiguity set $\Pi'$ is $N_D (\conv \Pi, r)$. If $N_D ( \conv \Pi,r) = \conv N_D ( \Pi, r)$, then the ambiguity sets $N_D ( \conv \Pi,r)$ and $N_D ( \Pi, r)$ determine the same maxmin objective. But in general, only the inclusion $\conv N_D ( \Pi, r) \subseteq N_D ( \conv \Pi,r)$ is guaranteed.} we show that the new mechanism obtained from this algorithm maximizes a particular variational preference relation in which the ambiguity index is proportional to the distance from $\Pi$.

\begin{prop}[New mechanism is a variational solution] \label{res:maxmin_variational_equivalence} Fix a feasible set $\FF$ and a nonempty, convex subset $\Pi$ of $\D(\Th)$. Consider a weak, convex metric $D$ on $\D(\Th)$ and a radius $r > 0$. Let $\hat{f}$ be a social choice function. If $\hat{f}$ is maxmin optimal with respect to the ambiguity set $N_D ( \Pi, r)$, then there exists $\l \geq 0$ such that
\begin{equation} \label{eq:variational}
    \hat{f} \in \argmax_{f \in \FF} \inf_{\pi \in \D( \Th)} \Brac{ \ang{u \circ f, \pi} + \l D(\pi, \Pi)}.
\end{equation}
\end{prop}

To prove \cref{res:maxmin_variational_equivalence}, we follow the proof of Lagrangian duality. We consider the problem of minimizing the function $\pi \mapsto \ang{ u \circ \hat{f}, \pi}$ subject to the constraint $D(\pi, \Pi) \leq r$. We take $\l$ to be the Lagrange multiplier on this constraint. Then we prove \eqref{eq:variational} using the maxmin optimality of $\hat{f}$ with respect to the ambiguity set $N_D(\Pi, r)$.\footnote{\citet[Section 5.2]{HansenEtal2006} prove a result similar to \cref{res:maxmin_variational_equivalence} in the case that $\Pi$ is a singleton and $D$ is relative entropy.} Note that the parameter $\l$ in \eqref{eq:variational} depends on the designer's feasible set $\FF$. \cref{res:maxmin_variational_equivalence} does not imply that the maxmin preferences (with respect to the metric neighborhood ambiguity set) coincide with the variational preferences induced by the representation in \eqref{eq:variational}. 


If $u$ is state-independent, then the objective $V(f) = \inf_{\pi \in \D( \Th)} [ \ang{u \circ f, \pi} + \l D(\pi,  \Pi)]$ from \eqref{eq:variational} is an example of the variational representation axiomatized by \citet{MaccheroniMarinacciRustichini2006}.\footnote{Similar to the maxmin case, this objective can be expressed as a minimum over all finitely additive priors; see \cref{sec:finitely_additive}.} For any social choice function $f$, the value $V(f)$ is the largest number $v$ such that
\[
    \ang{ u \circ f, \pi} \geq v - \l D( \pi, \Pi)\quad \text{for all}~\pi \in \D(\Th).
\]
For any prior $\pi$ in $\Pi$, we have $D(\pi, \Pi) = 0$, so the expected payoff from $f$ is guaranteed to be at least $V(f)$. At priors $\pi$ outside $\Pi$, the expected payoff from $f$ can drop below $V(f)$, but this deficit is controlled by the $D$-distance of $\pi$ from $\Pi$.

The objective $V(f)$ is similar to the special variational representation axiomatized by \cite{HansenEtal2024}. They distinguish ambiguity aversion from misspecification concerns. The set $\Pi$ is interpreted as a collection of ``structured models'' that captures the decision-maker's ambiguity. Their analog of the function $D$, which they require to be a statistical divergence, captures the decision-maker's concern that all of the structured models may be misspecified.\footnote{\cite{Lanzani2025} introduces a similar representation to capture misspecification concerns, but in his model, the decision-maker holds a prior over the set of structured models. Additionally, the agent dynamically adjusts his concern for misspecification in response to observed data.} Both our objective $V(f)$ and \citeapos{HansenEtal2024} objective give greater consideration to priors ``near'' the ambiguity set. The difference is in how ``nearness'' is quantified.  Our metric $D$ reflects the underlying topology on the state space. \cite{HansenEtal2024} do not consider a topology on the state space. In fact, their divergence is invariant to relabeling the states. Our metric $D$ is not a divergence in their sense because they require the divergence from $p$ to $q$ to be $\infty$ if $p$ is not absolutely continuous with respect to $q$. A leading example of a divergence is the relative entropy. With this choice of divergence and a singleton ambiguity set, the representation in \cite{HansenEtal2024} reduces to multiplier preferences \citep{HansenSargent2001}. 

To recover robustness, another reasonable approach is to work directly with the objective $V(f)$ rather than with the ambiguity set $N_D(\Pi, r)$. Our algorithm, however, illustrates that robustness can be recovered without leaving the class of maxmin preferences. In our algorithm, the primitive is the original ambiguity set $\Pi$. If the designer has a single prior $\pi_0$ in mind but uses a larger ambiguity set because of a concern that the prior $\pi_0$ is misspecified, then it is more natural to take $\Pi = \{ \pi_0\}$ in the algorithm.

Finally, we apply our algorithm to the regret-minimizing monopoly selling problem from \cite{BergemannSchlag2008}, described in 
\cref{sec:monopoly_selling_regret}. The state space is $\Th = [0, M]$, where we will take $M$ to be sufficiently large. The ambiguity set $\Pi$ contains all distributions that concentrate on the interval $[\ubar{\th}, 1]$, where the parameter $\ubar{\th}$ satisfies $0 \leq \ubar{\th} < 1$. We consider the usual absolute-value metric on $\Th$ and the associated Wasserstein metric $W$ on $\D(\Th)$. In the monopoly selling problem, every undominated mechanism corresponds to a distribution over posted prices. We represent a price distribution by its cumulative distribution function  $q \colon \R_+ \to [0,1]$. Thus, $q(\th)$ is the probability that type $\th$ buys the good. 

\begin{prop}[Robust regret-minimizing monopoly selling] \label{res:robustified_pricing_without_priors} Fix $r > 0$. Consider the regret-minimizing monopoly selling problem with $\Th = [0,M]$, where $M \geq e^{er}$. For each $\ubar{\th} \in [0,1)$, the following holds. 
\begin{enumerate}
    \item If $\ubar{\theta} \leq 1/e$, then the following mechanism is maxmin optimal with respect to the ambiguity set $N_W (\Pi, r)$:
    \begin{equation} \label{eq:q_BS}
        \hat{q}(\th) = \begin{cases} 
        0 &\text{if}~ \th < 1/e, \\
        1 + \ln \th &\text{if} ~1/e \leq \th < 1, \\
        1 &\text{if}~\th \geq 1.
        \end{cases}
    \end{equation}
    The associated regret guarantee is $1/e + r$.
    \item  If $\ubar{\th} > 1/e$, then the following mechanism is maxmin optimal with respect to the ambiguity set $N_W (\Pi, r)$:
    \begin{equation} \label{eq:q_revised}
    \hat{q}(\th) = 
    \begin{cases} 
        0 &\text{if}~ \th < \k( \ubar{\th},r),  \\
        \a( \ubar{\th}, r) \ln \frac{ \th}{\k ( \ubar{\th}, r)}
       &\text{if}~ \k (\ubar{\th}, r) \leq \th < \ubar{\th}, \\
        1 + \ln \th &\text{if}~ \ubar{\th} \leq \th < 1, \\
        1 &\text{if}~ \th \geq 1,
    \end{cases}
    \end{equation}
    where $\k(\ubar{\th}, r) = \ubar{\th} ( \ubar{\th} e)^{-1/ \a(\ubar{\th},r)} \in (1/e, \ubar{\th})$ and $\a (\ubar{\th}, r) \in [2, \infty)$; the function $\a$ is defined in the proof. The associated regret guarantee is 
    \[
        \ubar{\th} - \a ( \ubar{\th}, r) ( \ubar{\th} - \k ( \ubar{\th}, r)) + (\a (\ubar{\th}, r) - 1) r. 
    \]
\end{enumerate}
\end{prop}

\begin{figure}
\centering
\begin{tikzpicture}

    \begin{axis}[
        width=\textwidth,
        height = 0.6*\textwidth,
        xmin = 0.32, xmax = 1.1,
        ymin = 0.32, ymax = 0.52,
        axis lines = middle,
        xtick = {0.5, 1},
        xticklabels = {$\ubar{\th}$, $1$},
        ytick =  {0.346574, 0.5},
        yticklabels = {$R$, $\ubar{\th}$},
        xlabel = {$\th$},
        xlabel style={at=(current axis.right of origin), anchor=west},
        legend style={at={(0.85,0.7)}, anchor=south}
    ]

        \addplot[very thick, Blue,forget plot] coordinates {(0,0) (0.5,0.5)};
        \addplot[dashed, Blue,forget plot] (0.5,0.346574) -- (0.5,0.5);
        \filldraw[Blue] (0.5,0.346574) circle (2pt);
        \filldraw[Blue, fill=white] (0.5,0.5) circle (2pt);
        \addplot[very thick, Blue] coordinates {(0.5, 0.346574) (1,0.346574) (1.2, 0.546574)};
        
        \addplot[very thick, Orange] coordinates { (0,0) (0.468712,0.468712) (0.5, 0.351426) (1,0.351426) (1.2, 0.551426)};
        
        \addplot[very thick, Green] coordinates { (0,0) (0.446237,0.446237) (0.5, 0.354979)  (1,0.354979) (1.2, 0.554979)};

        \addplot[very thick, Purple] coordinates { (0,0) (0.428882,0.428882) (0.5, 0.357764) (1,0.357764) (1.2, 0.557764)};

        \legend{$r = 0$, $r = 0.001$, $r = 0.003$, $r= 0.006$};
        
    \end{axis}
\end{tikzpicture}
\caption{Robust regret-minimization with support restriction}
\label{fig:robustified_regret}
\end{figure}

We consider the two cases in turn. If $\ubar{\th} \leq 1/e$, then the maxmin-optimal mechanism from \cite{BergemannSchlag2008} coincides with $\hat{q}$ in \eqref{eq:q_BS}. This mechanism randomizes continuously over prices. The regret guarantee from this mechanism over $\Pi$ is robust, and the same mechanism remains maxmin optimal with respect to $N_W( \Pi, r)$ for each $r > 0$. If $\ubar{\th} > 1/e$, then the maxmin-optimal mechanism from \cite{BergemannSchlag2008} puts mass $1 + \ln \ubar{\th}$ on the price $p = \ubar{\th}$ and randomizes over prices $p$ in $[\ubar{\th}, 1]$ with density $1/p$. For each radius $r > 0$, the new maxmin-optimal mechanism with respect to $N_W (\Pi, r)$ retains the continuous part of the price distribution over $[\ubar{\th}, 1]$ but redistributes the point mass on $\ubar{\th}$ over the interval $[ \k (\ubar{\th}, r), \ubar{\th}]$, where the density is $\a ( \ubar{\th}, r) / p$. To see the significance of our method of enriching the ambiguity set, suppose instead that the analyst enlarged the ambiguity set by reducing the lower endpoint of the support from $\ubar{\th}$ to some $\ubar{\th}' \in (1/e, \ubar{\th})$. Then the associated maxmin-optimal mechanism would put mass $1 + \ln \ubar{\th}'$ on the price $p = \ubar{\th}'$. The performance of this mechanism drops discontinuously as the agent's valuation $\th$ falls below $\ubar{\th}'$.\footnote{Relatedly, \cite{Maselli2025} re-analyzes our monopoly selling problem with known median (\cref{sec:monopoly}) using a misspecification-averse objective of the form in \cite{HansenEtal2024}. The penalty function is a combination of the Wasserstein metric and the relative entropy.  Among all deterministic posted prices, the solution is a new, lower posted price.}

\cref{fig:robustified_regret} plots the regret functions induced by \citeapos{BergemannSchlag2008} original maxmin-optimal mechanism (labeled $r = 0$) and the new mechanisms (for three positive values of $r$), in the case $\ubar{\th} =  1/2$. The new mechanisms yield strictly lower regret than the original mechanism for valuations slightly below $\ubar{\th}$, at the cost of slightly higher regret over the interval $[\ubar{\th}, 1]$. As the radius $r$ increases, the optimal price distribution spreads out farther below $\ubar{\th}$, until $r$ reaches a critical radius $\hat{r}(\ubar{\th})$. In particular, $\hat{r} ( 1/2) \approx 0.0053$. Beyond this critical radius, the maxmin-optimal mechanism remains constant; the associated regret function, shown in purple, has the same absolute slope just below the left endpoint, $\ubar{\th}$, and just above the right endpoint, $1$. 

\section{Related literature} \label{sec:literature}

In this section, we relate our approach to other notions of robustness. It is helpful to distinguish between robustness to \emph{fundamentals} and robustness to \emph{information}.\footnote{For a survey of robustness in mechanism design, see \cite{Carroll2019}.} Our focus is on the robustness of the designer's payoff guarantee to small perturbations of the fundamentals. 

The paper most closely related to ours is \cite{MadaraszPrat2017}, which considers robustness to small perturbations of the fundamentals in a nonlinear pricing problem. The designer's ``model'' is a finite-support prior over the buyer's preferences. A mechanism that is optimal with respect to the designer's model can perform much worse if the preferences are slightly misspecified (with respect to the sup norm over the buyer's utility function). This fragility arises from binding non-local incentive constraints. We make a similar observation when analyzing saddle points with finite-support priors (\cref{res:saddle}), and we further show how worst-case priors with finite support can arise in maxmin problems. In their pricing problem, \cite{MadaraszPrat2017} modify the model-optimal mechanism so that it performs well under small preference misspecifications. While they consider the robustness of a particular Bayesian-optimal mechanism in a nonlinear pricing problem, we analyze the robustness of maxmin-optimal mechanisms in a general environment.

A distinctive feature of our analysis is that the notion of a ``small'' perturbation is determined by the topology on the state space. Few other papers in this literature explicitly use a topology on the state space.\footnote{The role of the state space topology is discussed in the literature on information costs. Standard information cost functions, such as entropy costs, are invariant to relabeling the states. This can yield unnatural consequences, as observed by \cite{PomattoStrackTamuz2023}. For instance, learning whether a country's GDP (to the nearest dollar) is above or below its median is just as costly as learning whether it is even or odd.} \cite{Stanca2023} provides an axiomatic foundation for a notion of robustness in a subjective expected utility setting, with a Polish state space and a Euclidean decision space. A menu containing continuous acts is robust if the decision-maker's value function from this menu is continuous in the prior (with respect to the weak topology). Under this definition, the decision-maker's choice from the menu varies as the prior changes. By contrast, we are interested in the designer's payoff guarantee from a fixed act (i.e., social choice function) when the priors in the ambiguity set are perturbed. The distinction between these two robustness notions is formalized by \cite{Prasad2003} in a Bayesian model. That paper also gives examples where the payoff from a fixed Bayesian optimal mechanism is sensitive to perturbations of the prior (in the weak topology).  

The classical literature on robustness to information aims to relax the implicit common-knowledge assumptions in multi-agent mechanism design; see \cite{BergemannMorris2013} for a survey. Our focus is instead on the robustness of the designer's payoff guarantee to the designer's misspecification of the distribution of fundamentals, primarily in single-agent settings. In our framework, the feasible set of social choice functions does not vary with the designer's prior over the state space. This framework can directly accommodate multi-agent design problems as long as the solution concept is independent of the agents' beliefs. For example, it is standard to use ex-post or dominant-strategy incentive compatibility; see, e.g., \cite{BeiEtal2019}, \cite{HeLi2022}, \cite{BachrachTalgam-Cohen2022}, and \cite{Zhang2022}.  

The literature on ``continuous implementation'' analyzes which social choice functions can be implemented in a way that is robust to small perturbations of the information structure \citep{OuryTercieux2012,JehielMeyerMoldovanu2012,ChenMuellerMallesh2022,ChenKunimotoSun2023}. In particular, \cite{Oury2015} and \cite{MeyerMorris2011} consider joint perturbations of information and fundamentals.\footnote{\cite{PeiStrulovici2024} introduce an ex-ante notion of local robustness. They consider perturbations in which agents' preferences can differ arbitrarily, with small probability. This notion does not use a topology on the space of preferences.} To capture informational perturbations in our framework, we could redefine the state to specify both the fundamental and the profile of agents' types, with some associated topology. With large type spaces, however, the appropriate topology is not clear and it would be cumbersome to specify a set of priors over this space.

Finally, a recent strand of the literature considers multi-agent design problems in which each mechanism is evaluated according to its worst payoff across all information structures and all associated equilibria. The designer has some knowledge of the payoff-relevant fundamentals---for example, the distribution or mean of the common value in an auction \citep{Du2018,BrooksDu2021a,BrooksDu2021b}, or the support of the sum of the agents’ valuations in a public-good problem \citep{BrooksDu2023}. A consistent finding of these papers is that it is optimal to use a mechanism in which bids are single-dimensional and each player's payment is \emph{proportional} to his bid. In our framework, we associate a payoff guarantee with each social choice function, rather than each mechanism. In our leading applications, this feasible set contains all social choice functions that can be induced by some choice of mechanism and some selection of an associated equilibrium. It is not clear how to capture adversarial equilibrium selection within this framework.

\section{Conclusion} \label{sec:conclusion}

In this paper, we introduce a robustness criterion for maxmin design problems. When solving for a maxmin-optimal mechanism, we propose that the analyst check whether the associated payoff guarantee is robust. We observe that many maxmin-optimal mechanisms from the literature fail our robustness criterion. This non-robustness often arises when the mechanism is obtained from a saddle point in which the prior has finite support. In this case, the associated maxmin-optimal mechanism is tailored to the particular states where this prior puts mass, making the mechanism simple but fragile. To recover robustness, we present an algorithm for enriching the ambiguity set to a metric neighborhood. The associated maxmin-optimal mechanism's payoff guarantee (over this new ambiguity set) is always robust.  

\newpage

\appendix

\section{Main proofs} \label{sec:main_proofs}

\subsection{Preliminaries}

All signed measures on $\Th$ are defined on the Borel $\s$-algebra. For any signed measure $\mu$ on $\Th$, define the total variation norm $\| \mu \|_{\mathrm{TV}}$ by 
\[
    \| \mu \|_{\mathrm{TV}} = \frac{1}{2} \sup_{\PP} \sum_{E \in \PP} |\mu (E)|,
\]
where the supremum is over all finite, measurable partitions $\PP$ of $\Th$. The factor of $1/2$ ensures that for $\pi, \pi' \in \D(\Th)$, we have $\| \pi - \pi' \|_{\mathrm{TV}}  = \sup_A | \pi (A) - \pi'(A)|$, where the supremum is over all Borel subsets $A$ of $\Th$.

By the Jordan decomposition theorem, each signed measure $\mu$ can be uniquely expressed as $\mu = \mu_+ - \mu_-$ for some nonnegative measures $\mu_+$ and $\mu_-$ that are mutually singular.\footnote{That is, there exists a Borel subset $A$ of $\Th$ such that $\mu_+ (A) = 0$ and $\mu_- (\Th \setminus A) = 0$.}  The \emph{support} of $\mu$, denoted $\supp \mu$,  is defined to be the support of the associated nonnegative measure $\mu_+ +\mu_-$.\footnote{The support of a nonnegative measure on a Polish space is the complement of the largest open set with measure $0$ (which can be shown to exist).} 

We extend the notation $\ang{ \cdot, \cdot}$ to integration against signed measures. Given a measurable function $v \colon \Th \to \R$ and a signed measure $\mu$ on $\Th$, define the integral of $v$ against $\mu$ by $\ang{v, \mu} = \ang{v, \mu_+} - \ang{ v, \mu_-}$, provided that $v$ is absolutely integrable with respect to $\mu_+$ and $\mu_-$. A signed measure $\mu$ on $\Th$ is \emph{finite} if $\| \mu \|_{\mathrm{TV}} < \infty$. A function $v \colon \Th \to \R$ is bounded if  $\| v\|_{\infty} \coloneqq \sup_{\th \in \Th} | v(\th)| < \infty$. For any bounded measurable function $v \colon \Th \to \R$ and any finite signed measure $\mu$, the integral $\ang{ v, \mu}$ is well-defined and satisfies $|\ang{ v, \mu}| \leq 2 \| v \|_{\infty} \| \mu \|_{\mathrm{TV}}$. Moreover, the map $\ang{ v, \cdot}$ is linear on the space of finite signed measures.  

Given a bounded function $v \colon \Th \to \R$, the \emph{lower semicontinuous envelope of $v$}, denoted $\lsc v$, is the pointwise-largest lower semicontinuous function that is pointwise smaller than $v$. It follows that the epigraph of $\lsc v$ is the closure of the epigraph of $v$ in the product topology on $\Th \times \R$.

\begin{lem}[Lower semicontinuous envelope and weak convergence]  \label{res:lsc_weak} Let $v \colon \Th \to \R$ be bounded and measurable. For each prior $\pi \in \D(\Th)$, there exists a sequence $(\pi_n)$ converging to $\pi$ such that 
$\lim_n \ang{ v, \pi_n} = \ang{\lsc v, \pi}$.
\end{lem}

The proofs of all lemmas are in \cref{sec:technical_proofs}.

\subsection{Countably additive versus finitely additive priors} \label{sec:finitely_additive}

Suppose that the utility function $u$ is state-independent.

First, we check that the payoff guarantee in \eqref{eq:objective} has a
\cite{GilboaSchmeidler1989} representation as a minimum over a set of finitely additive probability measures. In \cite{GilboaSchmeidler1989}, the primitive is a state space $\Th$ endowed with an algebra. We take the algebra to be the Borel $\s$-algebra $\BB(\Th)$. The space of finitely additive probability measures, $\D_{\operatorname{fa}}(\Th)$, is endowed with the topology of pointwise convergence on $\BB(\Th)$. The space $\D_{\operatorname{fa}}(\Th)$ is compact because it is a closed subset of $[0,1]^{\BB(\Th)}$, which is compact by Tychonoff’s theorem. For any social choice function $f$, the associated value function $u \circ f$ is measurable (by standard properties of probability kernels) and bounded. Therefore, $u \circ f$ can be uniformly approximated by measurable simple functions (i.e., functions with finite range). Thus, the extension of $\ang{ u \circ f, \cdot}$ to $\D_{\mathrm{fa}} (\Th)$ is well-defined, continuous, and affine. Let $\cconv \Pi$ denote the closed convex hull of $\Pi$ in $\D_{\mathrm{fa}} (\Th)$. We claim that
\[
    \inf_{ \pi \in \Pi} \ang{ u \circ f, \pi} = \min_{ \pi \in \cconv{\Pi}}  \ang{ u \circ f, \pi}.
\]
On the right, the minimum is achieved because $\cconv \Pi$ is compact and $\ang{ u \circ f, \cdot}$ is continuous. Equality holds because $\ang{ u \circ f, \cdot}$ is affine and continuous.

Next, we check that the variational objective in \eqref{eq:variational}  has a \cite{MaccheroniMarinacciRustichini2006} representation as a minimum over finitely additive priors. We prove a more general result. Consider a convex function $c \colon  \D(\Th) \to \R$ that is grounded (i.e., $\inf_{\pi \in \D(\Th)} c( \pi) = 0$). Let $\bar{c}$ be the extension of $c$ to the space $\D_{\mathrm{fa}} (\Th)$ that satisfies $\bar{c}( \pi) = \infty$ for all $\pi \in \D_{\mathrm{fa}}(\Th) \setminus \D(\Th)$. Its lower semicontinuous envelope, $\lsc \bar{c}$, is lower semicontinuous, convex, and grounded (i.e., $\inf_{\pi \in \D_{\mathrm{fa}}(\Th)} \lsc \bar{c}( \pi) = 0$). We claim that
\[
    \inf_{ \pi \in \D(\Th)} \Brac{ \ang{ u \circ f, \pi} + c ( \pi)} = \min_{ \pi \in \D_{\mathrm{fa}}(\Th)} \Brac{ \ang{ u \circ f, \pi} + \lsc \bar{c}  ( \pi)}.
\]
On the right, the minimum is achieved because the objective is lower semicontinuous and $\D_{\mathrm{fa}}(\Th)$ is compact. Equality holds because the function $\ang{ u \circ f, \cdot} + \lsc \bar{c}  ( \cdot)$ is the lower semicontinuous envelope of the function $\ang{ u \circ f, \cdot} + \bar{c}(\cdot)$ (since $\ang{u \circ f, \cdot}$ is continuous).

\subsection{Proof of Proposition~\ref{res:Hausdorff_lsc}} \label{sec:proof_Hausdorff_lsc}

Fix an ambiguity set $\Pi \in \KK$ and a social choice function $f \colon \Th \to \D(X)$. Define the function $W \colon \KK \to \R$ by $W(\Pi') = \inf_{\pi \in \Pi'} \ang{ u \circ f, \pi}$.

Suppose that the payoff guarantee from $f$ over $\Pi$ is not robust. Then there exists a sequence $(\pi_n)$ converging to a prior in $\Pi$ (which equals the closure of $\Pi$) such that 
\[
    \liminf_n  \ang{ u \circ f, \pi_n} < W(\Pi).
\]
For each $n$, let $\Pi_n = \Pi \cup \{\pi_n\}$. By construction, each $\Pi_n$ is compact, and the sequence $(\Pi_n)$ converges to $\Pi$ in the Hausdorff topology. We have
\[
    \liminf_n  W(\Pi_n) \leq    \liminf_n \ang{ u \circ f, \pi_n}  < W(\Pi),
\]
so $W$ is not lower semicontinuous at $\Pi$.

Conversely, suppose that the function $W$ is not lower semicontinuous at $\Pi$. Then there exists $\e > 0$ and a sequence $(\Pi_n)$ in $\mathcal{K}$ converging to $\Pi$ such that for each $n$, we have
\begin{equation*}
     W(\Pi_n) \leq  W(\Pi) - 2 \e. 
\end{equation*}
For each $n$, we can choose $\pi_n \in \Pi_n$ such that $\ang{u \circ f, \pi_n} \leq   W(\Pi_n) + \e$, hence $\ang{u \circ f, \pi_n} \leq W(\Pi) - \e$. To show that the payoff guarantee from $f$ over $\Pi$ is not robust, it suffices to find a subsequence of $(\pi_n)$ that converges to a prior in $\Pi$. Let $D$ be a metrization of the weak topology on $\D(\Th)$.\footnote{For example, let $D$ be the Prokhorov metric on $\D(\Th)$ associated with some compatible metric on $\Th$.} Since $(\Pi_n)$ converges to $\Pi$ in the Hausdorff metric associated with $D$ (see \cref{ft:metrization}), we can choose $\pi_n'$ in $\Pi$ for each $n$ such that $D(\pi_n, \pi_n') \to 0$. Since $\Pi$ is compact, the sequence $(\pi_n')$ has a subsequence that converges to some prior $\pi$ in $\Pi$. Hence, the corresponding subsequence of $(\pi_n)$ converges to $\pi$ as well. 

\subsection{Proof of Proposition~\ref{res:robustness_continuity}} \label{sec:proof_robustness_continuity}

To simplify notation, let $v_f = u \circ f$.

(i)  First, suppose that $v_f$ is continuous. Fix an ambiguity set $\Pi$. Let $(\pi_n)$ be a sequence in  $\D(\Th)$ that converges to some prior $\pi$ in the closure of $\Pi$. By the definition of the closure, we can choose a sequence $(\pi_n')$ in $\Pi$ that converges to $\pi$. By the portmanteau theorem, $\ang{ v_f, \cdot}$ is continuous on $\D(\Th)$, so 
\[
    \lim_n  \ang{v_f, \pi_n} = \ang{v_f,\pi} = \lim_n \ang{ v_f, \pi_n'} \geq \inf_{\pi' \in \Pi} \ang{v_f, \pi'}.
\]
Thus, the payoff guarantee from $f$ over $\Pi$ is robust.

Conversely, suppose that $v_f$ is not continuous. Then for some $\e > 0$, there exists a point $\th$ in $\Th$ and a sequence $(\th_n)$ converging to $\th$ such that either (a) $v_f(\th_n) \geq v_f(\th) + \e$ for all $n$, or (b) $v_f(\th_n) \leq v_f(\th) - \e$ for all $n$. If (a) holds, let $\Pi_1 = \{ \d_{\th_n} : n \geq 1 \}$. In this case, the payoff guarantee from $f$ over $\Pi_1$ is not robust because $\d_{\th}$ is in the closure of $\Pi_1$. If (b) holds, let $\Pi_2 = \{ \d_{\th} \}$. In this case, the payoff guarantee from $f$ over $\Pi_2$ is not robust because the sequence $(\d_{\th_n})$ converges to $\d_{\th}$.

(ii) First, suppose that $v_f$ is lower semicontinuous. Fix a closed ambiguity set $\Pi$.  Let $(\pi_n)$ be a sequence in $\D(\Th)$ that converges to some prior $\pi$ in the closure of $\Pi$. Since $\Pi$ is closed, $\pi$ is in $\Pi$. By the portmanteau theorem, $\ang{ v_f, \cdot}$ is lower semicontinuous on $\D(\Th)$, so
\[
    \liminf_n \ang{v_f, \pi_n} \geq \ang{v_f, \pi} \geq \inf_{\pi' \in \Pi} \ang{v_f, \pi'}.
\]
Thus, the payoff guarantee from $f$ over $\Pi$ is robust.

Conversely, suppose that $v_f$ is not lower semicontinuous. Then for some $\e > 0$, there exists a point $\th$ in $\Th$ and a sequence $(\th_n)$ converging to $\th$ such that  $v_f(\th_n) \leq v_f(\th) - \e$ for all $n$. The payoff guarantee from $f$ over the singleton $\{ \d_{\th} \}$ is not robust because the sequence $(\d_{\th_n})$ converges to $\d_{\th}$. Since the singleton $\{ \d_{\th} \}$ is closed, this completes the proof. 

(iii)--(iv) First, we prove (iv), which implies the `if' direction of (iii) since every singleton is closed. Fix a closed ambiguity set $\Pi$. Suppose that $\pi(D_- (v_f)) = 0$ for all priors $\pi \in \Pi$. Let $(\pi_n)$ be a sequence in $\D(\Th)$ that converges to some prior $\pi_0$ in the closure of $\Pi$.  Since $\Pi$ is closed, $\pi_0$  is in $\Pi$. By assumption, $\pi_0 ( D_- (v_f)) = 0$, so $\ang{ \lsc v_f, \pi_0} = \ang{v_f, \pi_0}$. Therefore,
\begin{equation*}
\begin{aligned}
    \liminf_n \ang{ v_f, \pi_n} 
   &\geq
    \liminf_n \ang{ \lsc v_f, \pi_n} \\
    &\geq 
    \ang{\lsc v_f, \pi_0} \\
    &=
     \ang{v_f, \pi_0} \\
     &\geq \inf_{\pi' \in \Pi} \ang{v_f, \pi'},
\end{aligned}
\end{equation*}
where the first inequality holds from the pointwise inequality $v_f \geq \lsc v_f$, and the second inequality follows from the portmanteau theorem. We conclude that the payoff guarantee from $f$ over $\Pi$ is robust.

Now we prove the `only if' direction of (iii). Fix $\pi$ in $\D(\Th)$. Suppose that $\pi (D_- (v_f)) > 0$. Hence, $\ang{ \lsc v_f, \pi} < \ang{v_f, \pi}$. By \cref{res:lsc_weak}, there exists a sequence $(\pi_n)$ converging to $\pi$ such that
\[
   \liminf_n \ang{v_f , \pi_n} \leq \ang{ \lsc v_f, \pi} <  \ang{v_f , \pi}.
\]
Thus, the payoff guarantee from $f$ over $\{\pi\}$ is not robust.

\subsection{Proof of Remark~\ref{res:AC}} \label{sec:AC_proof}

Consider the monopoly problem with $K$ goods, where $K \geq 1$.  
The state space is $\Th = [0, M]^K$ for some fixed $M > 0$. The state $\th = ( \th_1, \ldots, \th_K)$ denotes the agent's valuation for each of the $K$ goods. Since the players have additive, quasilinear utility, we can represent a decision lottery as a pair $(q,t) \in [0,1]^K \times \R$, where $q_k$ is the probability of receiving good $k$ and $t$ is the expected payment from the agent. For any incentive compatible social choice function $(q,t) \colon \Th \to [0,1]^K \times \R$, the designer's revenue in state $\th$ can be expressed as $\th^T q(\th) - U(\th)$, where $U$ is the agent's indirect utility function, which is convex and Lipschitz continuous. By incentive compatibility, the function $q$ is a selection from the subdifferential of $U$. The function $U$ is Lipschitz, so by Rademacher's theorem, $U$ is differentiable almost everywhere. It can be shown that $q$ is continuous wherever $U$ is differentiable,\footnote{Fix a differentiability point $\th$ and let $(\th_n)$ be a sequence converging to $\th$. For every $\th'$, we have $U(\th') \geq U(\th_n) + q(\th_n)^T (\th' - \th_n)$. Let $\bar{q}$ be a limit point of $(q(\th_n))$. Since $[0,1]^K$ is compact, it suffices to show that $\bar{q} = \nabla U(\th) = q(\th)$. Passing the limit along a subsequence, we conclude that $\bar{q}$ is a subgradient of $U$ at $\th$, hence $\bar{q} = q(\th)$.} so it follows that $q$ is continuous almost everywhere. We conclude that the designer's revenue from $(q,t)$  is continuous almost everywhere. The designer's regret from $(q,t)$ is $\th^T \mathbf{1} - (\th^T q(\th) - U (\th))$. This function is also continuous almost everywhere since $\th \mapsto \th^T \mathbf{1}$ is continuous everywhere. 

\subsection{Proof of Proposition~\ref{res:Bayesian_robustness}}

In the saddle point inequality \eqref{eq:saddle_def}, the first inequality is equivalent to Bayesian optimality of $\hat{f}$ under the prior $\hat{\pi}$, and the second inequality is equivalent to $\Pi \subseteq \Pi_0 ( \hat{f}, \hat{\pi})$. Finally, the requirement that $\hat{\pi}$ is in $\Pi$ is equivalent to $\{ \hat{\pi} \} \subseteq \Pi$.

Suppose that there does not exist $\th \in \Th$ such that $\hat{\pi} = \d_{\th}$ and $\{\th \} = \argmax_{\th' \in \Th} (u \circ \hat{f}) (\th')$. If $\hat{\pi}$ concentrates on $\argmax_{\th' \in \Th} (u \circ \hat{f}) (\th')$, then there exists $\th_0 \in \argmax_{\th' \in \Th} (u \circ \hat{f}) (\th')$ such that $\d_{\th_0} \neq \hat{\pi}$. If $\hat{\pi}$ does not concentrate on $\argmax_{\th' \in \Th} (u \circ \hat{f}) (\th')$, then we may select $\th_0 \in \Th$ such that $(u \circ \hat{f}) (\th_0) > \ang{ u \circ \hat{f}, \hat{\pi}}$. In both cases, $\d_{\th_0}$ is in $\Pi_0 (\hat{f}, \hat{\pi})$ and $\d_{\th_0} \neq \hat{\pi}$. In fact, since $\Pi_0 (\hat{f}, \hat{\pi})$ is convex, it follows that $[\d_{\th_0}, \hat{\pi}] \subseteq \Pi_0 (\hat{f}, \hat{\pi})$.

\subsection{Proof of Proposition~\ref{res:saddle}} \label{sec:saddle_proof}

Since the players have quasilinear utility, we can represent a decision lottery as a pair $(q,t) \in \D(Q) \times \R$, where $t$ is the \emph{expected} transfer. Linearly extend the functions $v$ and $c$ to $\D(Q)$. Let $(\hat{f}, \hat{\pi})$ be a saddle point. Write $\hat{f} = (\hat{q}, \hat{t})$. Since $\hat{f}$ is Bayesian optimal under $\hat{\pi}$, we know that $\hat{f}$ is not strictly dominated. Hence,  $\hat{t}$ is bounded.\footnote{Since $v$ is bounded, the participation constraint implies that $t$ is bounded above. Moreover, $t$ must be nonnegative; otherwise, the principal could translate the transfer function up by a constant $\e$ satisfying $0 < \e < - \inf_{\th \in \Th} t(\th)$. This modification preserves incentive compatibility (since the translation is the same for all reports) and participation (since every type can get nonnegative utility from making a report at which the new transfer is negative).} Suppose that $\hat{\pi}$ has finite support and $\ang{u \circ \hat{f}, \hat{\pi}} > 0$. We prove that the payoff guarantee from $\hat{f}$ over $\Pi$ is not robust. 
The proof uses two claims: 
 \begin{enumerate}
     \item There exists $\th_0 \in \supp \hat{\pi}$ such that $(u \circ \hat{f}) (\th_0) > 0$ and $v( \hat{q}(\th_0), \th_0) = \hat{t} (\th_0)$. 
     \item There exists a sequence $(\th_n)$ converging to $\th_0$ such that $\hat{t}(\th_n) = 0$ for each~$n$. 
 \end{enumerate}

We complete the proof, assuming these claims. For each $n$, let
\[
    \pi_n = \hat{\pi} + \hat{\pi} ( \th_0) ( \d_{\th_n} - \d_{\th_0}).
\]
Since the sequence $(\th_n)$ converges to $\th_0$, the sequence $(\pi_n)$ converges to $\hat{\pi}$.  For each $n$, we have
\begin{equation*}
\begin{aligned}
    \ang{ u \circ \hat{f}, \pi_n} 
    &=  \ang{u \circ \hat{f}, \hat{\pi}} + \hat{\pi} ( \th_0)[ (u \circ  \hat{f}) ( \th_n) - (u \circ  \hat{f}) ( \th_0)]  \\
    &\leq \ang{u \circ \hat{f}, \hat{\pi}} -  \hat{\pi} ( \th_0) (u \circ  \hat{f}) (\th_0) \\
    &= \inf_{\pi \in \Pi} \ang{ u \circ \hat{f}, \pi} - \hat{\pi} ( \th_0) (u \circ  \hat{f})(\th_0) \\
    &< \inf_{\pi \in \Pi} \ang{ u \circ \hat{f}, \pi},
\end{aligned}
\end{equation*}
where the inequalities follow from the claims (and the nonnegativity of the cost function $c$) and the second equality holds because $(\hat{f}, \hat{\pi})$ is a saddle point.

Now we verify the two claims. 
\begin{enumerate}
    \item  Let $\Th_0  = \{ \th \in \supp \hat{\pi} : (u \circ \hat{f})(\th) > 0 \}$. Since $\ang{u \circ \hat{f}, \hat{\pi}} > 0$, 
    the set $\Th_0$ is nonempty. From the participation constraint, we have $v( \hat{q}(\th), \th) \geq \hat{t}(\th)$ for all $\th$ in $\Th$. Suppose for a contradiction that $v(\hat{q}(\th), \th) > \hat{t}(\th)$ for all $\th$ in $\Th_0$. Let $\e = \min_{\th \in \Th_0} [ v(\hat{q}(\th),\th) - \hat{t}(\th)]$; this minimum exists because $\Th_0$ is finite. We construct a feasible social choice function $\tilde{f} = (\tilde{q}, \tilde{t})$ that the principal strictly prefers to $\hat{f}$. For each type $\th$ in $\Th_0$, let $(\tilde{q}(\th), \tilde{t}(\th)) = (\hat{q}(\th), \hat{t}(\th)+\e)$. For each type $\th \in \Th \setminus \Th_0$, let $(\tilde{q}(\th), \tilde{t}(\th))$ be an optimal choice for type $\th$ from the menu 
\[
\MM = \{ (\hat{q}(\th), \hat{t}(\th) + \e): \th \in \Th_0 \} \cup \{ (q_0, 0)\},
\]
using an arbitrary fixed tie-breaking rule. Since $\MM$ is finite, the map $(\tilde{q}, \tilde{t})$ is measurable. It is easily verified that $(\tilde{q}, \tilde{t})$ satisfies the incentive compatibility and participation constraints. For each $\th \in \Th_0$, we have $( u \circ \tilde{f}) (\th) = (u \circ \hat{f})(\th) + \e$. For each $\th \in \supp \hat{\pi} \setminus \Th_0$, we have $( u \circ \tilde{f}) (\th) \geq 0 \geq  (u \circ \hat{f})(\th)$. Therefore, $\ang{ u \circ \tilde{f}, \hat{\pi}}  - \ang{ u \circ \hat{f}, \hat{\pi}} \geq \e \hat{\pi} (\Th_0) > 0$, contrary to the optimality of $\hat{f}$ under $\hat{\pi}$.
    \item Let $\th_0$ be the type guaranteed by the first claim. By the fullness assumption, there exists a sequence $(\th_n)$ converging to $\th_0$ such that each type $\th_n$ is weaker than type $\th_0$. For each $n$, we have
\begin{equation*}
\begin{aligned}
    0 &= v(\hat{q}(\th_0), \th_0) - \hat{t}(\th_0)  \\
    &\geq v( \hat{q}(\th_n), \th_0) - \hat{t}(\th_n) \\
    &\geq v( \hat{q}(\th_n), \th_n) - \hat{t}(\th_n) \\
    &\geq 0,
\end{aligned}
\end{equation*}
where the first inequality follows from incentive compatibility, the second inequality holds because type $\th_n$ is weaker than type $\th_0$, and the last inequality follows from participation. Thus, all the inequalities hold with equality. In particular, $\hat{t}(\th_n) = v( \hat{q} (\th_n), \th_0) = v( \hat{q}(\th_n), \th_n)$. Type $\th_n$ is weaker than type $\th_0$, so we must have $v( \hat{q} (\th_n), \th_0) = 0$, hence $\hat{t}(\th_n) = 0$.
\end{enumerate}

\subsection{Proof of Proposition~\ref{res:rich}} \label{sec:proof_robust}

Let $\Pi$ be rich. We prove that $\Pi$ is uniformly robust. Fix a decision environment $(X, u)$. By definition, the function $u$ is bounded. Let $(\pi_n)$ be a sequence in $\D(\Th)$ that converges to a prior in the closure of $\Pi$. Since $\Pi$ is rich, there exists a sequence $(\pi_n')$ in $\Pi$ such that $\| \pi_n' - \pi_n \|_{\mathrm{TV}} \to 0$. For each social choice function $f \colon \Th \to \D(X)$, we have
\begin{equation*}
\begin{aligned}
    \ang{ u \circ f, \pi_n } 
    &= \ang {u \circ f, \pi_n'}  - \ang{ u \circ f,\pi_n' - \pi_n} \\
    &\geq \inf_{\pi \in \Pi} \ang{ u \circ f, \pi} - 2 \|u\|_{\infty} \| \pi_n' - \pi_n\|_\mathrm{TV}.
\end{aligned}
\end{equation*}
Rearranging and then passing to the infimum over all social choice functions $f$ gives
\[
\inf_{f} \Brac{ \ang{u \circ f, \pi_n} -  \inf_{\pi \in \Pi} \ang{ u \circ f, \pi} } \geq -2 \| u\|_{\infty} \| \pi_n' - \pi_n\|_{\operatorname{TV}}.
\]
Since $\| \pi_n' - \pi_n \|_{\mathrm{TV}} \to 0$, taking the limit infimum in $n$ gives the desired inequality. 

\subsection{Proof of Proposition~\ref{res:robust_sets}} \label{sec:proof_rich}

The proof is separated into parts. 

\paragraph{Preliminaries} First, we check that richness is implied by the following seemingly weaker property. An ambiguity set $\Pi$ is  \emph{approximately rich} if for every $\e > 0$, the following holds: for every sequence $(\pi_n)$ in  $\D(\Th)$ that converges to a prior in the closure of $\Pi$, there exists a sequence $(\pi_n')$ in $\Pi$ such that
\[
    \limsup_n \| \pi_n' - \pi_n \|_{\mathrm{TV}} \leq \e.
\]

Let $\Pi$ be an ambiguity set that is approximately rich. We prove that $\Pi$ is rich. Let $(\pi_n)$ be a sequence in $\D(\Th)$ that converges to a prior in the closure of $\Pi$. For each integer $j \geq 1$, take $\e = 1/j$ in the definition of approximate richness to conclude that there exists a sequence $(\pi_{j,n}')_{n=1}^{\infty}$ in $\Pi$ such that
\[
    \limsup_{n} \| \pi_{j,n}' - \pi_n \|_{\mathrm{TV}} \leq 1/j.
\]
Thus, for each $j$ there exists a positive integer $N_j$ such that
\[
    \sup_{n \geq N_j} \| \pi_{j,n}' - \pi_{n} \|_{\mathrm{TV}} \leq 2/j.
\]
Now we construct the desired sequence $(\pi_n')$ using a diagonalization argument. For $n < N_1$, let $\pi_n'$ be an arbitrary prior in $\Pi$. For $n \geq N_1$, let $\pi_n' = \pi_{j(n), n}'$, where $j(n)$ is the largest $j$ in $\{1, \ldots, n \}$ that satisfies $N_j \leq n$. By construction, for each $j \geq 1$, we have $\| \pi_n' - \pi_n \|_{\mathrm{TV}} \leq 2/j$ whenever $n \geq \max\{j, N_j\}$. We conclude that $\| \pi_n' - \pi_n \|_{\mathrm{TV}} \to 0$.

Second, we introduce the following result to deal with unbounded functions.
 
\begin{lem}[Unbounded moment approximation] \label{res:approximation} 
Let $(\pi_n)$ and $(\pi_n')$ be sequences in $\D(\Th)$ that weakly converge to the same prior $\pi$ in $\D(\Th)$. Let $H \colon \Th \to \R_+$ be continuous. For each $\e > 0$, there exists a sequence $(\rho_n)$ in $\D(\Th)$ weakly converging to $\pi$ such that
\begin{enumerate}[label = (\roman*), ref = \roman*]
    \item  $\| \rho_n - \pi_n \|_{\mathrm{TV}} \leq \e$ for each $n$;
    \item $H$ is bounded on $\cup_n \supp (\rho_n - \pi_n')$;
    \item for any continuous function $h \colon \Th \to \R$ satisfying $|h| \leq H$, we have $\ang{h, \rho_n - \pi_n'} \to 0$.
\end{enumerate}
\end{lem}

To prove \cref{res:approximation}, we construct each prior $\rho_n$ to agree with $\pi_n'$ when $H$ is sufficiently large and to approximate $\pi_n$ otherwise. 

To complete the main proof, we show that each form of ambiguity set is approximately rich. 

\paragraph{Continuous moment sets}
Let $\Pi = M(g,Y)$ for some continuous function $g \colon \Th \to \R^m$ and some star-$g$-interior subset $Y$ of $\R^m$. Let $(\pi_n)$ be a sequence in $\D(\Th)$ that converges to some prior $\pi$ in the closure of $\Pi$. Choose a sequence $(\pi_n')$ in $\Pi$ that converges to $\pi$. 

Let $H(\th) = \| g(\th) \|$, where $\| \cdot \|$ denotes the Euclidean norm on $\R^m$. Fix $\e > 0$. It follows from \cref{res:approximation} that there exists a sequence $(\rho_n)$ in $\D(\Th)$ converging to $\pi$ such that (i) $\limsup_n \| \rho_n - \pi_n \|_{\mathrm{TV}} \leq \e$; (ii) $H$ is bounded on $\supp (\rho_n - \pi_n')$ for each $n$; and (iii) $\lim_n \ang{g_j, \rho_n - \pi_n'} = 0$  for each $j =1, \ldots m$. Each prior $\pi_n'$ is in $\Pi$, so each component function $g_j$ is absolutely integrable with respect to $\pi_n'$ and hence, by (ii), also with respect to $\rho_n$. We conclude from (iii) that $\lim_n |\ang{ g_j, \rho_n} - \ang{g_j, \pi_n'} |  = 0$, for each $j =1, \ldots, m$. 

We construct a sequence $(\rho_n')$ in $\Pi$ by modifying the sequence $(\rho_n)$. Since $Y$ is star-$g$-interior, there exists $y_0$ in $\operatorname{relint} (\conv g(\Th))$ such that $[y_0, y] \subseteq Y \cap \conv g(\Th)$ for all $y \in  Y \cap \conv g(\Th)$. Let $B(c,r)$ (respectively, $\bar{B}(c,r)$) denote the open (respectively, closed) ball of radius $r$ with center $c$. Let  $\aff g(\Th)$ denote the affine hull of $g(\Th)$. Choose $\d >0$ such that $\bar{B}(y_0, \d) \cap \operatorname{aff} g(\Th) \subseteq \conv g(\Th)$. For each $n$, let $x_n = \ang{g, \rho_n}$  and  $y_n= \ang{g, \pi_n'}$. Since $\pi_n'$ is in $\Pi$, we know that $y_n$ is in $Y$. Also, $x_n$ and $y_n$ are both in $\conv g(\Th)$ since the expectation of a random vector lies in the convex hull of its range. Let\footnote{Here, we use the convention that $0/ \| 0 \| = 0$. Thus, $z_n = y_0$ if $x_n = y_n$.}
\begin{equation} \label{eq:replacement_1}
    z_n = y_0 + \d \frac{ y_n - x_n}{\| y_n - x_n \|} \in \bar{B} (y_0, \d) \cap \operatorname{aff} g(\Th) \subseteq \conv g(\Th).
\end{equation}
By Carath\'{e}odory's theorem, there exists a probability measure $\zeta_n$ in $\D(\Th)$ supported on at most $m +1$ points of $\Th$ such that $\ang{g, \zeta_n} =  z_n$. Let
\[
    \rho_n' = (1 - \a_n) \rho_n + \a_n \zeta_n, \quad\text{where}\quad \a_n = \frac{ \| y_n - x_n \|}{\d + \| y_n - x_n\|}.
\]
By the application of \cref{res:approximation} above, we have $ \| x_n - y_n\| \to 0$, so $\a_n \to 0$. For each $n$,\footnote{Here, we are extending the notation $\ang{ \cdot, \cdot}$ to integration of $\R^m$-valued functions.}
\begin{equation} \label{eq:replacement_2}
 \ang{g, \rho_n'}  = (1 - \a_n) x_n + \a_n z_n = (1 - \a_n) y_n + \a_n y_0 \in [y_0, y_n] \subseteq Y \cap \conv g(\Th),
\end{equation}
so $\rho_n'$ is in $\Pi$.  For each $n$, the triangle inequality gives 
\[
    \| \rho_n' - \pi_n\|_{\mathrm{TV}} \leq  \a_n + \| \rho_n - \pi_n\|_{\mathrm{TV}},
\]
so $\limsup_n \| \rho_n' - \pi_n \|_{\mathrm{TV}} \leq \e$, as desired. 

Since the finite union of rich ambiguity sets is rich,\footnote{This is immediate because the closure of a finite union equals the finite union of the closures.} we conclude that $M(g,Y)$ is rich if $g$ is continuous and $Y$ can be expressed as a finite union of star-$g$-interior sets. 

\paragraph{Metric neighborhoods} Let $\Pi = N_D(C, r)$ for some nonempty subset $C$ of $\D(\Th)$, some radius $r > 0$, and some weak, convex metric $D$ on $\D(\Th)$. Let $(\pi_n)$ be a sequence in $\D(\Th)$ that converges to some prior $\pi$ in the closure of $\Pi$. By the definition of the closure, there exists a sequence $(\pi_n')$ in $\Pi$ that converges to $\pi$.  Since $D$ induces the weak topology on $\D(\Th)$, we have $D(\pi_n', \pi_n) \to 0$. By the definition of $\Pi$, there exists a sequence $(\z_n)$ in $C$ such that $\limsup_n D( \z_n, \pi_n') \leq r$. Therefore, 
\begin{equation} \label{eq:alpha_lim}
    \limsup_{n \to \infty} D(\z_n, \pi_n) \leq r.
\end{equation}
For each $n$, let
\[
    \rho_n' = (1 - \a_n) \pi_n + \a_n \z_n, 
    \quad
    \text{where}
    \quad
    \a_n = (1 - r /D (\z_n, \pi_n))_+, 
\]
with the convention that $\a_n = 0$ whenever $D(\z_n, \pi_n) =0$. By \eqref{eq:alpha_lim}, $\a_n \to 0$, hence $\| \rho_n' - \pi_n \|_{\mathrm{TV}} \to 0$. We claim that, for each $n$,  the prior $\rho_n'$ is in $\Pi$. By the  convexity of the metric $D$,
\begin{equation*}
    D ( \rho_n', \z_n) 
    \leq   (1 - \a_n) D( \pi_n, \z_n) 
    \leq r,
\end{equation*}
where the second inequality follows from the definition of $\a_n$, upon separating into cases according to whether $D(\pi_n, \z_n) \leq r$. Since $\z_n$ is in $C$, it follows that $\rho_n'$ is in $\Pi$, as needed.\footnote{The same proof goes through if $\Pi$ is an \emph{open} neighborhood. Simply replace $\a_n$ with $\min \{1, \a_n + 1/n\}$. We still get $\a_n \to 0$, and now $D ( \rho_n', \z_n) < r$.}

\paragraph{Wasserstein distance with unbounded metric} We modify the above proof to cover the case that $D$ is the Wasserstein distance $W$ with respect to some unbounded, compatible metric $d$ on $\Th$. In this case, weak convergence does not imply convergence in $W$. 
To complete the proof, consider the sequences $(\pi_n)$, $(\pi_n')$, $(\z_n)$  and the prior $\pi$ as above. Choose $\th_0$ in $\Th$ and let $H(\th) = d (\th_0, \th)$. Fix $\e > 0$.  By \cref{res:approximation}, there exists a sequence $(\rho_n)$ in $\D(\Th)$ converging to $\pi$ such that (i) $\limsup_n \| \rho_n - \pi_n \|_{\mathrm{TV}} \leq \e$, and (ii) $H$ is bounded on $\cup_n \supp (\rho_n - \pi_n')$. Since $(\pi_n')$ and $(\rho_n)$ both converge to $\pi$, it can be shown using (ii) that $W(\rho_n, \pi_n') \to 0$.\footnote{We adopt the notation from \citet[pp.~109--110]{Bogachev2018}. For any bounded signed measure $\mu$ on $\Th$ with $\mu(\Th) = 0$, let $\| \mu \|_{\mathrm{KR}} = \sup_f \ang{f, \mu}$, where the supremum is over all $1$-Lipschitz functions $f \colon \Th \to \R$ with $\| f\|_{\infty} \leq 1$. If additionally $\supp \mu$ is $d$-bounded, let $\| \mu \|_{\mathrm{K}} = \sup_f \ang{f, \mu}$, where the supremum is over all $1$-Lipschitz functions $f \colon \Th \to \R$. Let $L = \sup_{\th \in \cup_n  \supp (\rho_n - \pi_n')} H(\th)$. By  \citet[3.2.7 Theorem, p.~114]{Bogachev2018}, we have 
\[
     W ( \rho_n, \pi_n')
    = \| \rho_n  - \pi_n' \|_{\mathrm{K}} 
    \leq \max\{ 2 L, 1\} \| \rho_n  - \pi_n' \|_{\mathrm{KR}}.
\]
Finally, since $\| \cdot \|_{\mathrm{KR}}$  metrizes the weak topology on $\D(\Th)$ \citep[3.2.2 Theorem, p.~111]{Bogachev2018}, we have $\| \rho_n - \pi_n'\|_{\mathrm{KR}} \to 0$.}
Therefore, we can follow the rest of the proof as above, with $(\rho_n)$ in place of $(\pi_n)$, to construct a sequence $(\rho_n')$ in $\Pi$ satisfying $\| \rho_n'  - \rho_n \|_{\mathrm{TV}} \to 0$. By (i), we conclude that $\limsup_n \| \rho_n' - \pi_n \|_{\mathrm{TV}} \leq \e$. 

\subsection{Proof of Proposition~\ref{res:maxmin_variational_equivalence}}

Since $\Pi$ is convex and $D$ is a convex function on $\D(\Th) \times \D(\Th)$, it is easily verified that the function $D(\cdot, \Pi)$ is convex. For the first part of the proof, we apply a separating hyperplane argument, essentially following the proof of strong Lagrangian duality under Slater's condition.

Let $\a = \inf_{\pi \in N_D (\Pi, r)} \ang{ u \circ \hat{f}, \pi}$.  Consider the sets
\begin{equation*}
      A = \set{ \paren{ D(\pi, \Pi), \ang{u \circ \hat{f}, \pi} } : \pi \in \D( \Th)} + \R_+^2, \qquad
      B = [0, r] \times (-\infty, \a).
\end{equation*}
It is easily verified that both sets are convex. By construction, $A$ and $B$ are disjoint. By the separating hyperplane theorem, there exists a nonzero vector $v = (v_1, v_2)$ satisfying $v^Ta \geq v^T b$ for all $a \in A$ and $b \in B$. Since $A$ is upward closed, we must have $v \geq 0$.  Since $\Pi$ is nonempty, the set $A$ contains a point with first coordinate $0$. The set $B$ contains a point with first coordinate $r$, which is strictly positive. Therefore, $v_2 > 0$. After renormalizing, we may assume that $v = ( \l, 1)$ for some $\l \geq 0$. Thus, 
\[
    \inf_{\pi \in \D(\Th)} \Brac{   \ang{ u \circ \hat{f}, \pi} + \l D( \pi, \Pi)} \geq \a +  \l r.
\]
On the other hand, for each $f$ in $\FF$, we have
\begin{equation*}
\begin{aligned}
    \inf_{\pi \in \D( \Th)} \Brac{ \ang{u \circ f, \pi} + \l D (\pi, \Pi)} 
    &\leq 
    \inf_{\pi \in N_D (\Pi, r)} \Brac{ \ang{u \circ f, \pi} + \l D (\pi, \Pi)} \\
    &\leq   \inf_{\pi \in N_D (\Pi, r)} \Brac{ \ang{u \circ f, \pi}} + \l r \\
    &\leq \a + \l r,
\end{aligned}
\end{equation*}
where the last inequality holds because $\hat{f}$ is maxmin-optimal with respect to $N_D( \Pi, r)$. We conclude that $\hat{f} \in \argmax_{f \in \FF} \inf_{\pi \in \D(\Th)} [ \ang{ u \circ f, \pi} + \l D(\pi, \Pi)]$. 

\subsection{Proof of Proposition~\ref{res:robustified_pricing_without_priors}}

Fix $r > 0$. Let $\Th = [0, M]$ for some $M$ satisfying $M \geq e^{e r}$. 

By the envelope theorem, a quantity function $q \colon \Th \to [0,1]$ together with a transfer function $t \colon \Th \to \R$ satisfies incentive compatibility and participation if and only if $q$ is weakly increasing, $t(0) \leq 0$, and
\begin{equation} \label{eq:envelope}
    \th q(\th) - t(\th) =  - t(0) + \int_{0}^{\th} q(\th') \de \th' \quad\text{for all}~ \th \in \Th.
\end{equation}
Every incentive compatible mechanism is weakly dominated by some incentive compatible mechanism in which $q$ is right-continuous, $t(0) = 0$, and $q(M) = 1$. Therefore, we may restrict attention to mechanisms satisfying these conditions. Any such mechanism can be interpreted as a random (nonnegative) price, with cumulative distribution function $q$. The associated regret function is given by
\begin{equation} \label{eq:regret_formula_twice}
    R( \th ; q) = \th -\int_{0}^{\th} \th' \de q(\th') = \th (1 - q(\th)) + \int_{0}^{\th} q(\th') \de \th'.
\end{equation}

Fix $\ubar{\th} \in [0,1)$. Let $\Pi = \{ \pi \in \D( \Th): \pi ([\ubar{\th}, 1]) = 1\}$. Let $\Pi' = N_W(\Pi,r)$. For any $\pi \in \D(\Th)$, the Wasserstein distance from $\pi$ to $\Pi$ can be expressed as
\[
    W(\pi, \Pi) = \int_{0}^{\ubar{\th}} (\ubar{\th} - \th) \de \pi (\th) + \int_{1}^{M} (\th - 1) \de \pi (\th).
\]
We will repeatedly use this formula below. We separate the proof into cases. 

\paragraph{Case 1} Suppose $\ubar{\th} \leq 1/e$.  We construct a saddle point. Let $\hat{q}$ be the price distribution defined in \eqref{eq:q_BS}. Define the distribution $\hat{\pi} \in \D(\Th)$ by the cumulative distribution function
\[
    F_{\hat{\pi}}(\th)
    =
    \begin{cases}
        0 &\text{if } \th < 1/e,\\[3pt]
        1 - \dfrac{1}{e\th} &\text{if } 1/e \leq \th < e^{er} ,\\[8pt]
        1 &\text{if } \th \geq e^{er}.
    \end{cases}
\]
It is easily verified that $W(\hat{\pi},\Pi)=r$.\footnote{Since $\ubar{\th}\le 1/e$, only the part of $\hat{\pi}$ above $1$ contributes to $W(\hat{\pi}, \Pi)$. We have
\[
    W (\hat{\pi}, \Pi) =  \int_0^{e^{er} - 1} (1 - F_{\hat{\pi}}(\th + 1)) \de \th = \frac{1}{e} \int_0^{e^{er} - 1} \frac{1}{\th+ 1} \de \th  = r.
\]
} We check that $(\hat{q},\hat{\pi})$ is a saddle point.
\begin{itemize}
    \item First, we check that $\hat{q}$ is a best response to $\hat{\pi}$. Under the prior $\hat{\pi}$, the designer's expected revenue from posting price $p \geq 0$ is
\[
    p(1-F_{\hat{\pi}}(p-))
    =
    \begin{cases}
        p &\text{if } p \leq 1/e,\\[3pt]
        1/e &\text{if } 1/e < p \leq e^{er},\\[3pt]
        0 &\text{if } p > e^{er}.
    \end{cases}
\]
Since the price distribution $\hat{q}$ concentrates on $[1/e,1]$, we conclude that $\hat{q}$ is a best response to $\hat{\pi}$.\footnote{Under a fixed prior, a mechanism maximizes expected revenue if and only if it minimizes expected regret.}

    \item Next, we check that $\hat{\pi}$ is a best response, over $N_W(\Pi,r)$, to $\hat{q}$. The regret function induced by $\hat{q}$ can be expressed as
\[
    R(\th;\hat{q}) = 1/e + (\th - 1)_+ -  (1/e - \th)_+.
\]
In particular, $R(\th;\hat{q}) = 1/e$ for all $\th$ in $[1/e,1]$. For every $\pi \in N_W ( \Pi, r)$, the expected regret from $\hat{q}$ is at most $1/e + r$, with equality if $\pi$ concentrates on $[1/e, M]$ and satisfies $W( \pi, \Pi) = r$. Thus, $\hat{\pi}$ is a best response to $\hat{q}$.
\end{itemize}

\paragraph{Case 2} Suppose $\ubar{\th} > 1/e$. We begin by defining the function $\a$ appearing in \eqref{eq:q_revised}. Let
\begin{equation} \label{eq:radius_def}
    \hat{r}(\ubar{\th}) = 
    \sqrt{ \ubar{\th} /e} \int_{ \sqrt{\ubar{\th}/e}}^{\ubar{\th}}  \frac{ \ubar{\th} - \th}{\th^2}  \de \th =   \ubar{\th}  -  (1/2) \sqrt{\ubar{\th} /e } \Paren{ 3 + \ln \ubar{\th}}.
\end{equation}
For $r \geq \hat{r} (\ubar{\th})$, let $\a(\ubar{\th}, r) = 2$. For $r < \hat{r}(\ubar{\th})$, define $\a(\ubar{\th}, r) \in (2, \infty)$ implicitly by 
\begin{equation} \label{eq:alpha_def}
    r = \ubar{\th}( \ubar{\th} e)^{-1/
        \a(\ubar{\th},r)} \int_{\ubar{\th}( \ubar{\th} e)^{-1/
        \a(\ubar{\th},r)}}^{\ubar{\th}}  \frac{\ubar{\th} - \th}{\th^2} \de \th.
\end{equation}
It can be verified that $\a$ is well-defined, and $\a$ is strictly decreasing in $r$ with $\lim_{r \uparrow \hat{r}(\ubar{\th})} \a (\ubar{\th},r) = 2$.\footnote{Evaluating the integral in \eqref{eq:alpha_def}, we see that $\a ( \ubar{\th}, r)$ is defined by $r = \psi (\ubar{\th}, \a(\ubar{\th}, r))$, where
\[
    \psi ( \ubar{\th}, \a) =   \ubar{\th} -  \a^{-1} \ubar{\th} (\ubar{\th} e)^{-1 / \a} ( \a + 1 + \ln \ubar{\th}).
\] Since $\ubar{\th} > 1/e$, the expression $\psi ( \ubar{\th}, \a)$ is strictly decreasing in $\a$ over $(0, \infty)$, with $\psi ( \ubar{\th}, 2) = \hat{r}(\ubar{\th})$ and $\lim_{\a \to \infty} \psi ( \ubar{\th}, \a) = 0$; to see this, observe that the partial derivative 
\[
    D_2 \psi ( \ubar{\th}, \a) = - \a^{-3} e^{-1/\a} \ubar{\th}^{1 - 1/\a} (1 + \ln \ubar{\th})^2
\]
is strictly negative.}  Recall the definition $\k ( \ubar{\th}, r) = \ubar{\th} ( \ubar{\th} e)^{-1 / \a( \ubar{\th}, r)}$. Since $\ubar{\th} > 1/e$ and $\a ( \ubar{\th}, r) \geq 2$, we have $1/e < \k (\ubar{\th},r) < \ubar{\th}$. 

Next, let
\begin{equation} \label{eq:beta_def}
    \b(\ubar{\th},r) = \exp \Set{ (r - \hat{r}(\ubar{\th}))_+ \sqrt{e/\ubar{\th}} }.
\end{equation}
If $r < \hat{r} (\ubar{\th})$, then  $\b(\ubar{\th},r) = 1$. If $r \geq \hat{r} (\ubar{\th})$, then since $\ubar{\th} > 1/e$, we have $\b(\ubar{\th},r) < e^{er}$.

With these definitions in place, we now construct a saddle point. Let $\hat{q}$ be the price distribution defined in \eqref{eq:q_revised}. Define the distribution $\hat{\pi} \in \D(\Th)$ by the cumulative distribution function 
\[
    F_{\hat{\pi}}(\th)  
    =
     \begin{cases}
        0 &\text{if}~ \th < \k (\ubar{\th},r), \\
        1 - \frac{\k (\ubar{\th},r)}{\th}  &\text{if}~\k (\ubar{\th},r) \leq \th < \b(\ubar{\th}, r),  \\
        1 &\text{if}~ \th \geq \b(\ubar{\th}, r).
    \end{cases}
\]
If $r < \hat{r}(\ubar{\th})$, then $\b ( \ubar{\th}, r) = 1$, and  the definition of $\a ( \ubar{\th},r)$ in \eqref{eq:alpha_def} ensures that $W(\hat{\pi}, \Pi) = r$.\footnote{Over the range $[ \k (\ubar{\th}, r), \ubar{\th}]$, the distribution $\hat{\pi}$ has density $\k ( \ubar{\th}, r)/ \th^2$, so the integral on the right side of \eqref{eq:alpha_def} equals $W( \hat{\pi}, \Pi)$.} If $r \geq \hat{r} ( \ubar{\th})$, then the definitions of $\hat{r} (\ubar{\th})$ in \eqref{eq:radius_def} and $\b( \ubar{\th},r)$ in \eqref{eq:beta_def} ensure that $W(\hat{\pi}, \Pi) = r$.\footnote{By \eqref{eq:radius_def}, the part of $\hat{\pi}$ below $\ubar{\th}$ contributes $\hat{r} (\ubar{\th})$ to $W( \hat{\pi}, \Pi)$. The part of $\hat{\pi}$ above $1$ contributes
\[
    \int_0^{\b ( \ubar{\th}, r)- 1} (1 - F_{\hat{\pi}}(\th + 1)) \de \th =  \sqrt{ \ubar{\th} /e} \int_0^{\b ( \ubar{\th}, r) - 1} \frac{1}{\th+ 1} \de \th  = \sqrt{ \ubar{\th}/e} \ln \b ( \ubar{\th}, r),
\]
which equals $r - \hat{r} ( \ubar{\th})$ by 
\eqref{eq:beta_def}.} We check that $(\hat{q}, \hat{\pi})$ is a saddle point. 
\begin{itemize}
    \item First, we check that $\hat{q}$ is a best response to $\hat{\pi}$. Under the prior $\hat{\pi}$, the designer's expected revenue from posting price $p \geq 0$ is
\[
    p (1 - F_{\hat{\pi}}(p-)) 
    =
    \begin{cases} 
    p &\text{if}~p \leq \k ( \ubar{\th}, r), \\
   \k (\ubar{\th}, r) &\text{if}~ \k (\ubar{\th}, r) < p \leq \b(\ubar{\th}, r), \\
        0 &\text{if}~ p > \b (\ubar{\th}, r).
    \end{cases}
\]
Since the price distribution $\hat{q}$ concentrates on $[\k (\ubar{\th},r), 1]$, we conclude that $\hat{q}$ is a best response to $\hat{\pi}$.

\item Next, we check that $\hat{\pi}$ is a best response, over $N_W (\Pi, r)$, to $\hat{q}$. The regret function induced by $\hat{q}$ can be expressed as
\[
   R(\th ; \hat{q}) = R_0
    + ( \th -1)_+ 
    + (\a (\ubar{\th}, r) - 1) ( \ubar{\th} - \th)_+
    - \a (\ubar{\th}, r) \Paren{ \k (\ubar{\th},r) - \th}_+,
\]
where
\[
    R_0 = \k (\ubar{\th},r)- (\a (\ubar{\th}, r) - 1)(\ubar{\th} - \k (\ubar{\th},r)).
\]
In particular, $R(\th; \hat{q}) = R_0$ for all $\th$ in $[\ubar{\th}, 1]$. We separate into cases. 
\begin{enumerate}
    \item Suppose $r < \hat{r}(\ubar{\th})$. Then $\a(\ubar{\th}, r) - 1 > 1$. In this case, for every $\pi \in N_W ( \Pi, r)$, the expected regret from $\hat{q}$ is at most $R_0 + ( \a(\ubar{\th}, r) - 1 ) r$, with equality if $\pi$ concentrates on $[\k (\ubar{\th},r), 1]$. Thus, $\hat{\pi}$ is a best response to $\hat{q}$.
    \item Suppose $r \geq  \hat{r}(\ubar{\th})$. Then $\a(\ubar{\th}, r) - 1 = 1$. In this case, for every $\pi \in N_W ( \Pi, r)$, the expected regret from $\hat{q}$ is at most $R_0 + r$, with equality if $\pi$ concentrates on $[\k (\ubar{\th},r), M]$. Thus, $\hat{\pi}$ is a best response to $\hat{q}$.
\end{enumerate}
\end{itemize}

\newpage

\bibliography{lit.bib}
\bibliographystyle{ecta.bst}

\newpage

\section{Online appendix: Additional results} \label{sec:alternative_axiomatizations}

\subsection{Robustness of the payoff guarantee in \texorpdfstring{\cite{Carroll2015}}{Carroll~(2015)}}
\label{sec:linear_contracts}

In the robust contracting setting of \cite{Carroll2015}, there exists a maxmin-optimal contract that is linear. We show that the principal's payoff guarantee from any \emph{nonzero} maxmin-optimal linear contract is robust.

To make our robustness result stronger, we consider a larger state space, which allows for additional output and cost perturbations. Recall that the output set $\YY$ in \cite{Carroll2015} is a compact subset of $\R_+$ satisfying $\min \YY = 0$. Let $\YY'$ be a compact subset of $\R$ satisfying $\YY' \supseteq \YY$. In particular, $\YY'$ can contain negative output levels. A \emph{technology} is a nonempty compact subset of $\D( \YY') \times \R$. Define the state space $\Th$ to be the collection of all technologies. 

We now define a metric on $\Th$. First, endow $\YY'$ with the usual absolute value metric, and let $d_W$ denote the associated Wasserstein metric on $\D( \YY')$. Next, define the metric $d$ on $\D(\YY') \times \R$ by 
\begin{equation} \label{eq:define_d}
    d( (F_1, c_1), (F_2, c_2) ) = d_W (F_1, F_2) + | c_1 - c_2|.
\end{equation}
Finally, let $d_H$ denote the Hausdorff metric on $\Th$ associated with the metric $d$. 

With this larger state space, a \emph{contract} is a continuous function $w \colon \YY' \to \R_+$. Any linear contract on $\YY$ can be extended to $\YY'$ by paying nothing if the output is negative. Formally, for each $\a \in [0,1]$, define $w_\a \colon \YY' \to \R_+$ by $w_\a ( y) = \a y_+$. We will refer to $w_\a$ as a linear contract, even though it is not technically a linear function over its entire domain. 

We adopt the notation from \cite{Carroll2015}. Given a contract $w \colon \YY' \to \R_+$ and a technology $\AA$, the agent's expected utility is given by
\[
    V_A (w |\AA) = \max_{(F,c) \in \AA} \Paren{ \E_{F} [ w(y)] - c}.
\]
Let $\AA^\ast (w | \AA) = \argmax_{(F,c) \in \AA} \Paren{ \E_{F} [ w(y)] - c}$. With ties broken in the principal's favor, the principal's expected profit is given by
\[
    V_{P} (w | \AA) = \max_{(F,c) \in \AA^\ast (w | \AA)}  \E_{F} [ y - w(y)].
\]

We consider the same ambiguity set as in \cite{Carroll2015}. Let $\AA_0$ be a nonempty compact subset of $\D(\YY) \times \R_+$ satisfying the following nontriviality assumption: there exists $(F,c)$ in $\AA_0$ such that $\E_F[y]  > c$.  Let $\SS = \SS (\AA_0, \YY)$ be the collection of all technologies $\AA$ satisfying $\AA_0 \subseteq \AA \subseteq \D (\YY ) \times \R_+$. The principal's payoff guarantee from contract $w$ is given by
\[
    V_P (w) = \inf_{\AA \in \SS} V_P (w |\AA).
\]
Equivalently, 
\[
    V_P (w) = \inf_{\pi \in \Pi} \ang{ V_P (w |\cdot), \pi},
\]
where $\Pi = \Pi ( \AA_0, \YY)$ is the ambiguity set containing all priors in $\D(\Th)$ that concentrate on $\SS$. To be sure, technologies in $\SS$ contain actions only in $\D(\YY) \times \R_+$. But with our definition of $\Th$, the robustness definition considers perturbed technologies containing actions in $\D(\YY') \times \R$.

By \citet[Theorem 1, p.~541]{Carroll2015}, there exists a linear contract that maximizes $V_P$. To analyze the robustness of the associated payoff guarantee, we derive a lower bound on the principal's payoff from a given nonzero linear contract under each technology $\AA$.

\begin{prop}[Linear contract payoff bound]  \label{res:linear_contract_bound} Fix $\a \in (0,1]$ and let $L(\a) = \max \{ 1/\a , 2 (1 - \a)/\a\}$. For every technology $\AA$, we have
\begin{equation} \label{eq:lower_bound}
    V_P ( w_\a | \AA) \geq \frac{ 1- \a}{\a} V_A ( w_\a | \AA_0) - L(\a) \inf_{\AA' \in \SS} d_H ( \AA, \AA'). 
\end{equation}
\end{prop}

\begin{proof}  For any number $z$, let $z_{\pm} = \max \{ \pm z, 0\}$. For each $y \in \YY'$, we have
\begin{equation} \label{eq:pointwise}
    y  - w_\a( y) =  y_+ - y_- -  \a y_+  = \frac{1 - \a}{\a} w_\a (y) - y_-.
\end{equation}
Fix a technology $\AA$. Let $(F,c)$ be an action in $\AA^\ast ( w_\a | \AA)$. Taking expectations in \eqref{eq:pointwise} with respect to $F$ gives
\begin{equation} \label{eq:profit}
\begin{aligned}
     V_P (w_\a |\AA) 
     &\geq \E_F [ y - w_\a (y)] \\
     &= \frac{1 - \a}{\a} \E_F [ w_\a (y)] - \E_F [y_-] \\
     &= \frac{1- \a}{\a} (V_A ( w_\a | \AA) + c) - \E_F[y_-] \\
     &\geq  \frac{1- \a}{\a} V_A ( w_\a | \AA) - \max\{ (1 - \a)/\a, 1\} (\E_{F} [y_-] +  c_-).
\end{aligned}
\end{equation}

The map $(F',c') \mapsto \E_{F'} [w_\a(y)] - c'$ on $\D(\YY') \times \R$ is $1$-Lipschitz with respect to the metric $d$ from \eqref{eq:define_d}. Therefore, for every $\AA' \in \SS$, we have
\begin{equation*}
\begin{aligned}
    V_A( w_\a |  \AA) 
    &\geq V_A (w_\a | \AA') -  d_H (\AA, \AA') \\
    &\geq V_A (w_\a | \AA_0) -  d_H (\AA, \AA').
\end{aligned}
\end{equation*}
Taking the supremum over all $\AA'$ in $\SS$ gives
\begin{equation} \label{eq:util_bound}
  V_A( w_\a |  \AA)  \geq  V_A (w_\a | \AA_0) - \inf_{\AA' \in \SS} d_H (\AA, \AA'). 
\end{equation}
Also, it can be verified that\footnote{Fix $\AA'$ in $\SS$. For each $(F', c')$ in $\AA'$, we have $F' \in \D(\YY) \subseteq \D(\R_+)$ and $c' \geq 0$, so
\[
    d( (F,c), (F',c'))  = d_W( F,F') + |c - c'| \geq \E_F [ y_-] + c_-.
\]
Since $(F,c)$ is in $\AA$, it follows that $d_H (\AA, \AA') \geq \E_F [ y_-] + c_-$. Taking the infimum over all such $\AA'$ in $\SS$ yields the desired inequality.}
\begin{equation} \label{eq:y_bound}
\E_{F} [y_-] + c_- \leq \inf_{\AA' \in \SS} d_H (\AA, \AA').
\end{equation}
To complete the proof, substitute \eqref{eq:util_bound} and \eqref{eq:y_bound} into \eqref{eq:profit}, noting that
\[
\frac{1- \a}{\a}  + \max \Set{ \frac{1 - \a}{\a},1} = L(\a). \qedhere 
\]
\end{proof}

Recall from \citet[p.~542]{Carroll2015} that a contract $w$ is \emph{eligible} if $V_P (w) \geq V_P(w_0)$ and $V_P(w) > 0$. The nontriviality assumption on $\AA_0$ ensures that the optimal payoff guarantee is strictly positive, hence every maxmin-optimal contract is eligible.

\begin{prop}[Robustness of payoff guarantee]  \label{res:robust_guarantee} Fix $\a \in (0,1]$. If $w_\a$ is eligible, then the payoff guarantee from $w_\a$ over $\Pi$ is robust.
\end{prop}

\begin{proof} Suppose that $w_\a$ is eligible. 
Define $h_\a \colon \Th \to \R$ by 
\[
    h_\a(\AA) = \max \Set{  \frac{ 1- \a}{\a} V_A ( w_\a | \AA_0) - L(\a) \inf_{\AA' \in \SS} d_H ( \AA, \AA'), \min \YY' }.
\]
The function $h_\a$ is continuous and bounded. Recall that $\min \YY' \leq 0$. For every $y \in \YY'$, we have $y - w_\a(y) \geq \min \YY'$, so $V_P (w_\a | \AA )\geq \min \YY'$ for every technology $\AA$. By \cref{res:linear_contract_bound}, we conclude that $V_P (w_\a | \AA) \geq h_\a(\AA)$ for every technology $\AA$. Since $\min \YY' \leq 0$, for each $\AA \in \SS$ we have
\begin{equation} \label{eq:V_P}
h_\a (\AA) = \frac{ 1- \a}{\a} V_A ( w_\a | \AA_0)  = V_P (w_\a),
\end{equation}
where the second equality holds because $w_\a$ is eligible \citep[Lemma~2, pp.~542--543]{Carroll2015}.

Let $(\pi_n)$ be a sequence in $\D(\Th)$ that converges to some prior $\pi$ in the closure of $\Pi$. By the definition of the closure, we can choose a sequence $(\pi_n')$ in $\Pi$ that converges to $\pi$. By the portmanteau theorem, 
\[
    \liminf_n  \ang{ V_P (w_\a | \cdot), \pi_n} \geq  \lim_n \ang{ h_\a, \pi_n} = \ang{ h_\a, \pi} = \lim_n \ang{h_\a, \pi_n'} = V_P(w_\a),
\]
where the last equality follows from \eqref{eq:V_P}.
\end{proof}

\begin{rem}[Zero contract] If the zero contract is eligible, then its payoff guarantee is not robust. To see this, consider for each $\e > 0$ the 
technology 
\[
    \AA_\e = \{ (F,c+\e): (F,c) \in \AA_0 \} \cup \{ (\d_0,0) \}.
\]
Since $d_H ( \AA_\e, \AA_0 \cup \{ (\d_0, 0)\}) \leq \e$, it follows that $\inf_{\AA' \in \SS} d_H (\AA_\e, \AA') \leq \e$. Given the zero contract, action $(\d_0,0)$ is the agent's unique optimal choice from $\AA_\e$. Therefore, $V_P (w_0 | \AA_\e) = 0$. But $V_P( w_0) > 0$ since $w_0$ is eligible. 
\end{rem}

Suppose that $\AA_0$ does not contain a zero-cost action that generates strictly positive expected output. In this case, the calculations in \citet[p.~545]{Carroll2015} imply that the zero contract is \emph{not} maxmin optimal. Hence, \cref{res:robust_guarantee} implies that the payoff guarantee from every maxmin-optimal linear contract is robust.

\subsection{Continuity of payoff guarantee in the radius} \label{sec:continuity_in_radius}

Here, we show that the payoff guarantee from each social choice function is continuous in the radius used in our algorithm from \cref{sec:algorithm}.

\begin{prop}[Continuity in radius] \label{res:continuity_radius} Fix a feasible set $\FF$. Let $\Pi$ be an ambiguity set and let $D$ be a convex metric on $\D(\Th)$. For each $f$ in $\FF$, define the function $V_f \colon (0, \infty) \to \R$ by $V_f (r) = \inf_{\pi \in N_D (\Pi, r)} \ang{ u \circ f, \pi}$. Define $V \colon (0, \infty) \to \R$ by $V(r)= \sup_{f \in \FF} V_f(r)$. 
\begin{enumerate}[label = (\roman*)]
    \item The family $\{ V_f \}_{f \in \FF}$ is pointwise equicontinuous, hence $V$ is continuous. 
    \item If $\Pi$ is convex, then $V_f$ is convex for each $f$ in $\FF$, and hence $V$ is convex. 
\end{enumerate}
\end{prop}

\begin{proof} We prove each part separately. 
\begin{enumerate}[label = (\roman*)]
    \item  Fix $f \in \FF$ and $r  > r' > 0$. Fix $\e > 0$ and choose $\pi$ in $N_D( \Pi, r)$ such that $\ang{ u \circ f, \pi} \leq V_f(r) + \e$. Since $\pi$ is in $N_D( \Pi, r)$, we can choose $\pi_0 \in \Pi$ such that $D(\pi_0, \pi) \leq r + \e$. Let $\l = r'/ (r + \e)$ and $\pi' = \l \pi + (1 - \l) \pi_0$. Since $D$ is convex, we have
    \[
        D ( \pi', \pi_0) \leq  \l D (\pi, \pi_0) \leq r'.
    \]
    Thus, $\pi'$ is in $N_D ( \Pi, r')$. Therefore, 
    \begin{equation*}
    \begin{aligned}
        V_f (r') 
        &\leq \ang{ u \circ f, \pi'} \\ 
        &= \l \ang{u \circ f, \pi} + (1 - \l) \ang{u \circ f, \pi_0} \\
        &= \ang{ u \circ f, \pi} + (1 - \l) \Brac{\ang{u \circ f, \pi_0} - \ang{u \circ f, \pi} } \\
        &\leq V_f ( r) + \e + (1 - \l) 2 \| u \|_{\infty} \\
        &= V_f ( r) + \e + \frac{r + \e - r'}{r + \e} 2 \| u \|_{\infty}. 
    \end{aligned}
    \end{equation*}
    Since $\e$ was arbitrary, we can pass to the limit as $\e \to 0$ to conclude that 
    \begin{equation} \label{eq:direction_1}
    \begin{aligned}
      V_f ( r') \leq V_f(r) + \frac{2 \| u \|_{\infty}}{r} (r - r').
    \end{aligned}
    \end{equation}
    The function $V_f$ is weakly decreasing, so
    \begin{equation} \label{eq:equicont}
        | V_f(r') - V_f(r) | \leq \frac{2 \| u\|_{\infty}}{r} |r - r'|.
    \end{equation}
    This bound \eqref{eq:equicont} holds for each radius $r >0$ and each social choice function $f \in \FF$. Therefore, the family $\{ V_f\}_{f \in \FF}$ is pointwise equicontinuous. It follows that $V$ is continuous. 

    \item Suppose that $\Pi$ is convex. Fix $f \in \FF$.  Fix $r_1, r_2 > 0$. Fix $\e > 0$. For $i =1,2$, choose $\pi_i \in N_D ( \Pi, r_i)$ such that $\ang{ u \circ f, \pi_i} \leq V_f(r_i) + \e$. It can be verified that $D (\cdot, \Pi)$ is a convex function (because the metric $D$ is a convex function and $\Pi$ is a convex set). Therefore, for each $\l$ in $(0,1)$, we have 
    \begin{equation*}
    \begin{aligned}
        D(\l \pi_1 + (1 - \l) \pi_2, \Pi) 
        &\leq \l D(\pi_1, \Pi)  + (1 - \l) D (\pi_2, \Pi)\\
        &\leq \l r_1 + (1 - \l) r_2.
    \end{aligned}
    \end{equation*}
    Thus, $\l \pi_1 + (1 - \l) \pi_2$ is in $N_D(\Pi, \l r_1 + (1 - \l) r_2)$, and we conclude that 
    \begin{equation*}
    \begin{aligned}
        V_f ( \l r_1 + (1 - \l) r_2) 
        &\leq \ang{ u \circ f, \l \pi_1 + (1 - \l) \pi_2 } \\
        &= \l \ang{u \circ f, \pi_1} + (1 - \l) \ang{ u \circ f, \pi_2} \\
        &\leq \l V_f ( r_1) + (1 - \l)  V_f(r_2) + \e. 
    \end{aligned}
    \end{equation*}
    Since $\e$ is arbitrary, we conclude that $V_f$ is convex. Therefore, the function $V$ is the pointwise supremum of convex functions, and hence is convex.   \qedhere
\end{enumerate}
\end{proof}

\subsection{Ambiguity sets that are not uniformly robust} \label{sec:not_globally_robust}

We show that other commonly used classes of ambiguity sets are not uniformly robust. 

\begin{itemize}

    \item Suppose that $\Th$ is a convex subset of $\R$. For any $\pi$ in $\D(\Th)$ and $\a \in [0,1]$, let $Q_\a (\pi)$ denote the set of $\a$-quantiles of $\pi$.\footnote{That is, $Q_\a( \pi)$ contains all $x$ for which  $\pi (-\infty, x] \geq \a$ and $\pi [x, \infty) \geq 1-  \a$.}  A \emph{quantile set} is defined by
    \[
        Q ( (x_j, \a_j)_{j=1}^m) = \{ \pi \in \D(\Th): x_j \in Q_{\a_j} (\pi)~\text{for all}~j= 1, \ldots, m\},
    \]
    for some positive integer $m$ and some $x_1, \ldots, x_m \in \R$ and $\a_1, \ldots, \a_m \in [0,1]$ satisfying $\inf \Th < x_1 < \cdots < x_m < \sup \Th$ and $\a_1 < \cdots < \a_m$.\footnote{Here, $\inf \Th \in [-\infty, \infty)$ and $\sup \Th \in (-\infty, \infty]$.} The monopoly pricing example in \cref{sec:fragile} uses an ambiguity set of this form. 
    
    \item A \emph{probability set} is defined by 
    \[
      P(A,\a, \b) = \{ \pi \in \D(\Th): \a \leq \pi (A) \leq \b \},
    \]
    for some measurable proper subset $A$ of $\Th$ and some $\a, \b \in [0,1]$ with $\a \leq \b$. In particular, $P(A,1,1)$ is the \emph{support set} that contains all priors $\pi$ with $\pi (A) = 1$. 

    \item A \emph{support--moment set} is defined by 
  \[
        M (S;g,Y) = \Set{ \pi \in \D(\Th):\pi (S)= 1~\text{and}~ \E_{\th \sim \pi} [g(\th)] \in Y},
    \]
    for some measurable proper subset $S$ of $\Th$, some continuous function $g \colon \Th \to \R^m$, and some subset $Y$ of $\R^m$ that intersects the relative interior of $\conv g(S)$.

    \item Given priors $\mu, \nu \in \D(\Th)$, write $\mu \ll \nu$ if $\mu$ is absolutely continuous with respect to $\nu$. The \emph{relative entropy} (Kullback--Leibler divergence) is defined by
    \[
        R ( \mu \parallel \nu) 
        =
        \begin{cases}
            \ang{  \ln ( \frac{\de \mu}{\de \nu}), \mu} &\text{if}~ \mu \ll \nu, \\
            \infty &\text{otherwise},
        \end{cases}
    \]
    where $\frac{\de \mu}{\de \nu}$ denotes the Radon--Nikodym derivative of $\mu$ with respect to $\nu$. The \emph{relative entropy ball} of radius $r > 0$ about the reference prior $\nu$, denoted $B_R (\nu, r)$, contains all priors $\mu$ satisfying $R( \mu \parallel \nu) \leq r$. Relative entropy balls are used in \emph{constraint preferences} \citep{HansenSargent2001}. Total variation balls, denoted by $B_{\mathrm{TV}} (\nu, r)$, are defined analogously using the total variation norm.

    \item Suppose that $\Th = \prod_{j=1}^{k} \Th_j$, for some Polish spaces $\Th_1, \ldots, \Th_k$. A \emph{marginal set} is defined by
    \[
        \Gamma ( (\pi_j)_{j \in J}) = \{ \pi \in \D ( \Th) : \marg_j \pi = \pi_j~\text{for all}~j \in J\},
    \]
    for some nonempty subset $J$ of $\{1, \ldots, k\}$ and some probability measures $\pi_j \in \D(\Th_j)$ for each $j$ in $J$. \cite{Carroll2017} uses an ambiguity set of this form; see \cref{sec:Carroll}.
\end{itemize}

Some topological assumptions on the state space $\Th$ are needed to show that these ambiguity sets are not uniformly robust. In particular, we must rule out the discrete topology on $\Th$, for otherwise all ambiguity sets are uniformly robust. The space $\Th$ is \emph{perfect} if it has no isolated points. The space $\Th$ is \emph{connected} if it cannot be expressed as a disjoint union of two nonempty open sets. If $\Th$ is connected and has at least two points, then $\Th$ is perfect.

\begin{prop}[Not uniformly robust] \label{res:non_robust_sets}
Nonempty, proper subsets of $\D(\Th)$ taking the following forms are not uniformly robust:
\begin{enumerate}[label = (\roman*), ref = \roman*]
\item quantile sets, provided that $\Th$ is a convex subset of $\R$;
\item probability sets and support--moment sets, provided that $\Th$ is connected;
\item singletons,  relative entropy balls, and total variation balls, provided that $\Th$ is perfect;
\item marginal sets, provided that $\Th$ is a product of perfect spaces.
\end{enumerate}
\end{prop}

The proof is in \cref{sec:proof_non_robust}.

\subsection{Proof of Proposition~\ref{res:non_robust_sets}} \label{sec:proof_non_robust}

In this proof, we use the notation $W_\Pi (v)$ to denote $\inf_{\pi \in \Pi} \ang{v, \pi}$, for any ambiguity set $\Pi$ and any bounded, measurable function $v \colon \Th \to \R$. 

To prove that an ambiguity set $\Pi$ is not uniformly robust, it suffices to find a bounded, measurable function $v \colon \Th \to \R$ and a sequence $(\pi_n)$ converging to a prior in the closure of $\Pi$ such that  $\liminf_n \ang{ v, \pi_n} < W_\Pi (v)$. Then it is straightforward to construct a decision environment $(X,u)$ and a social choice function $f \colon \Th \to \D (X)$ such that $u \circ f  = v$.\footnote{Indeed, let $X = v(\Th)$ and let $u(x,\th) = x$. Then let $f(\th) = \d_{v(\th)}$.}
 
We organize the proof into parts according to the form of the ambiguity set. Some parts use the following approximation result. For any Borel subset $S$ of $\Th$, we view $\D(S)$ as a subset of $\D(\Th)$. 

\begin{lem}[Concentrated approximation] \label{res:concentration} Suppose that $\Th$ is perfect. For any prior $\pi$ in $\D(\Th)$ there exist a Borel subset $A$ of $\Th$ with $\pi(A) =1$ and a sequence $(\pi_n)$ in $\D(\Th \setminus A)$ that weakly converges to $\pi$.
\end{lem}

\paragraph{Quantile sets} Let $\Th$ be a convex subset of $\R$. Let $\Pi = Q ( (x_j, \a_j)_{j=1}^{m})$, for some positive integer $m$ and some $x_1, \ldots, x_m \in \Th$ and $\a_1, \ldots, \a_m \in [0,1]$ satisfying $\inf \Th < x_1 < \cdots < x_m < \sup \Th$ and $\a_1 < \cdots < \a_m$. If $\a_m = 0$, then $m =1$ and $\Pi$ is a support set, which is a special case of a probability set, which is considered below. Therefore, we may assume $\a_m > 0$.  Set $\a_0 = 0$. Let
\[
    \pi = (1 - \a_{m-1}) \d_{x_m} 
 + \sum_{j =1}^{m-1} (\a_j - \a_{j-1}) \d_{x_j}.
\]
By construction, $\pi$ is in $\Pi$.  Set $v = 1_{\Th \cap (-\infty, x_m]}$. We have $W_\Pi (v) = \a_m$. Choose a strictly decreasing sequence $(\th_n)$ in $\Th$ that converges to $x_m$. For each $n$, let $\pi_n = \pi + (1 - \a_{m-1})(\d_{\th_n} - \d_{x_m})$. The sequence $(\pi_n)$ converges to $\pi$, but $\ang{v, \pi_n} = \a_{m-1} < \a_m$ for all $n$.

\paragraph{Probability sets} Suppose that $\Th$ is connected. Let $\Pi = P(A,\a, \b)$ for some measurable proper subset $A$ of $\Th$ and some $\a,\b \in [0,1]$ with $\a \leq \b$. Suppose that $\Pi$ is a nonempty proper subset of $\D( \Th)$. It follows that  $[\a, \b] \neq [0,1]$. We may assume without loss that $\a > 0$; otherwise, we must have $\b < 1$, and we can express $P(A, \a, \b)$ as $P(A^c, 1 - \b, 1 - \a)$. Let $v = 1_A$. Thus, $W_\Pi (v) = \a$. Since $\Th$ is connected, $A$ cannot be both open and closed. We consider two (overlapping) cases.

First,  suppose that $A$ is not open. Choose $\th \in A \setminus A^\circ$, where $A^\circ$ denotes the interior of $A$. Choose a sequence $(\th_n)$ in $\Th \setminus A$ that converges to $\th$. Fix $\th' \in \Th \setminus A$. For each $n$, let $\pi_n =  \a \d_{\th_n} + (1 - \a) \d_{\th'}$. The sequence $(\pi_n)$ converges to $\pi = \a \d_{\th} + (1 - \a) \d_{\th'} \in \Pi$, but for all $n$, we have $\ang{v, \pi_n} = 0 < \a$.

Second, suppose that $A$ is not closed. Choose $\th \in \bar{A} \setminus A$, where $\bar{A}$ denotes the closure of $A$. Choose a sequence $(\th_n)$ in $A$ that converges to $\th$. For each $n$, let $\pi_n = \a \d_{\th_n} + (1 - \a ) \d_{\th} \in \Pi$. The sequence $(\pi_n)$ converges to $\d_{\th}$, so $\d_{\th}$ is in the closure of $\Pi$. Since $\ang{v, \d_\th} = 0 < \a$, the constant sequence $(\d_{\th})$ gives the desired inequality.

\paragraph{Support--moment sets}
Suppose that $\Th$ is connected.  Let $\Pi = M (S;g, Y)$ for some measurable proper subset $S$ of $\Th$, some continuous function $g \colon \Th \to \R^m$, and some subset $Y$ of $\R^m$ that intersects the relative interior of $\conv g(S)$. Choose $y_0 \in Y \cap \operatorname{relint} (\conv g(S))$.  It can be shown that there exists $\d > 0$ such that (a) $D \coloneq \bar{B}(y_0, \d) \cap \operatorname{aff} g(S) \subseteq \conv g(S)$, and (b) there exists a continuous function $\tilde{g} \colon D \to \D(S)$ such that $\ang{g, \tilde{g} (z)}  = z$, for each $z$ in $D$.\footnote{By the definition of the relative interior, we may choose $\d' >0$ such that $B(y_0, \d') \cap \operatorname{aff} g(S) \subseteq \conv g(S)$. Let $k-1$ denote the linear dimension of $\operatorname{aff} g(S)$. Choose affinely independent vectors $z_1, \ldots, z_k \in B(y_0, \d') \cap \operatorname{aff} g(S)$ so that $y_0$ is in $\operatorname{relint} (\conv (z_1, \ldots, z_k))$. Thus, there exists $\d$ in $(0, \d')$ such that 
\[
   D \coloneq \bar{B}(y_0, \d) \cap \operatorname{aff} g(S) \subseteq \conv (z_1, \ldots, z_k) \subseteq  B(y_0, \d') \cap \operatorname{aff} g(S). 
\]
Each vector $z_j$ is in $\conv g(S)$, so by Carath\'{e}odory's theorem, we may select a probability measure $\zeta_j$ in $\D(S)$ concentrating on at most $m +1$ points such that $\ang{g, \zeta_j} = z_j$. Let $\hat{p} = (\hat{p}_1, \ldots, \hat{p}_k)$ be the barycentric-coordinate mapping from $\conv (z_1, \ldots, z_k)$ to the probability simplex in $\R^k$. Thus, $z = \sum_{j=1}^{k} \hat{p}_j(z) z_j$, for each $z$ in $\conv (z_1, \ldots, z_k)$. For each $z$ in $D$, let $\tilde{g}(z) = \sum_{j=1}^{k} \hat{p}_j(z) \zeta_j$. By linearity,
$\ang{g, \tilde{g}(z)} = \sum_{j=1}^{k} \hat{p}_j(z) z_j = z$. The function $\tilde{g} \colon D \to \D(S)$ is continuous because the barycentric-coordinate map $\hat{p}$ is continuous.} Let $\bar{S}$ denote the closure of $S$. Since $g$ is continuous, we have $\aff g(S) = \aff g( \bar{S})$. Using $\tilde{g}$, we define a map $\tilde{\pi} \colon \bar{S} \to \D(\Th)$ as follows. For each  $\th$ in $\bar{S}$, let\footnote{Here, we adopt the convention that $0/\|0\|= 0$. If $g(\th) = y_0$, then $z(\th) = y_0$.}
\[
        z (\th) = y_0 + \d \frac{ y_0 - g(\th)}{\| y_0 -g(\th) \| } \in D,
\]
and let 
\[
        \tilde{\pi} (\th) = (1 -  \a(\th)) \d_{\th} +  \a (\th)  \tilde{g}(z(\th)),
        \quad
        \text{where}
        \quad
        \a (\th) = \frac{ \| y_0 - g(\th)\| }{\d + \| y_0 - g(\th)\|}.
    \]
For each $\th$ in $\bar{S}$, we have
\[
    \ang{g, \tilde{\pi} (\th)} = (1 - \a(\th)) g(\th) + \a(\th) z(\th) =  y_0 \in Y.
\]
Thus, $\tilde{\pi} (\th)$ is in $\Pi$ if $\th$ is in $S$. The map $\tilde{\pi}$ is continuous because $g$ and $\tilde{g}$ are continuous and $\a(\th_n) \to 0$ for any sequence $(\th_n)$ with $g(\th_n) \to y_0$.

We now complete the proof. Let $v = 1_S$. We have $W_\Pi (v) = 1$. Since $\Th$ is connected, $S$ cannot be both open and closed. We consider two (overlapping) cases. 

First, suppose that $S$ is not open. Choose a sequence $(\th_n)$ in $\Th \setminus S$ that converges to some point $\th$ in $S$. Let 
\begin{equation*}
\begin{aligned}
 \pi  &= (1 - \a(\th)) \d_{\th} + \a (\th) \tilde{g}(z(\th)),  \\
   \pi_n &= (1 - \a(\th)) \d_{\th_n} + \a (\th) \tilde{g}(z(\th)),
\end{aligned}
\end{equation*}
for each $n$.  Note that $\pi = \tilde{\pi} (\th)$. Since $\th$ is in $S$, we know that $\pi$ is in $\Pi$. The sequence $(\pi_n)$ converges to $\pi$, but for all $n$, we have $\ang{v, \pi_n} = \a(\th) < 1$. 

Second, suppose that $S$ is not closed. Choose $\th \in \bar{S} \setminus S$. Choose a sequence $(\th_n)$ in $S$ that converges to $\th$. For each $n$, let $\pi_n = \tilde{\pi} (\th_n)$ and let $\pi = \tilde{\pi} ( \th)$. Since $(\th_n)$ is in $S$, the sequence $(\pi_n)$ is in $\Pi$.  Since $\tilde{\pi}$ is continuous, the sequence $(\pi_n)$ converges to $\pi$, so $\pi$ is in the closure of $\Pi$. Since $\ang{v, \pi} = \a(\th) < 1$, the constant sequence $(\pi)$ gives the desired inequality.

\paragraph{Singletons, relative-entropy balls, and total-variation balls} Suppose that $\Th$ is perfect. Fix $\pi_0$ in $\D(\Th)$. We consider ambiguity sets of the following three forms: (a) $\Pi = \{ \pi_0 \}$; (b) $\Pi = B_R ( \pi_0, \b)$ for some $\b > 0$; and (c) $\Pi = B_{\mathrm{TV}} (\pi_0, \g)$ for some $\g \in (0,1)$. By \cref{res:concentration}, there exists a set $A$ with $\pi_0 (A) = 1$ and a sequence $(\pi_n)$ in $\D(\Th \setminus A)$ that converges to $\pi_0$. Let $v = 1_{A}$. For each $n$, we have $\ang{ v, \pi_n} = 0$. On the other hand, $W_\Pi (v) = 1$ in cases (a) and (b). In case (c), $W_\Pi (v) \geq 1- \g$ because for each $\pi' \in \Pi$, 
\[
   \ang{v, \pi'} = \pi' (A) \geq  \pi_0(A) - \| \pi' - \pi_0 \|_{\mathrm{TV}} \geq 1 - \g.
\]

\paragraph{Marginal sets} Let $\Th = \prod_{j=1}^{m} \Th_j$, where $\Th_1, \ldots, \Th_m$ are perfect. Let $\Pi = \G ( (\pi_j)_{j \in J})$ for some nonempty subset $J$ of $\{1, \ldots, m\}$ and some priors $\pi_j$ in $\D(\Th_j)$ for each $j$ in $J$. Without loss, suppose $1$ is in $J$. Apply \cref{res:concentration} at the prior $\pi_1$ in $\D(\Th_1)$ to obtain a subset $A_1$ of $\Th_1$ with $\pi_1 ( A_1) = 1$ and a sequence $(\pi_1^n)$ in $\D(\Th_1 \setminus A_1)$ that converges to $\pi_1$. For $j \notin J$, arbitrarily choose $\pi_j \in \D(\Th_j)$. For each $n$, let $\pi^n = \pi_1^n \otimes \pi_{-1}$, where $\pi_{-1} = \otimes_{j \neq 1} \pi_j$. The sequence $(\pi^n)$ converges to $\pi_1 \otimes \pi_{-1} \in \Pi$. Let $v = 1_{ A_1 \times \Th_{-1}}$. Thus, $W_\Pi (v) = 1$, but for each $n$, we have $\ang{v, \pi^n} = 0$.\footnote{For the case $m = 1$, interpret $\pi_1^n \otimes \pi_{-1}$ and $\pi_1 \otimes \pi_{-1}$ as $\pi_1^n$ and $\pi_1$, respectively.}

\newpage

\section{Online appendix: Proofs of technical lemmas} \label{sec:technical_proofs}

\subsection{Proof of Lemma~\ref{res:lsc_weak}}

Fix a bounded, measurable function $v \colon \Th \to \R$ and a prior $\pi$ in $\D(\Th)$. For any sequence $(\pi_n)$ converging to $\pi$, we have
\[
    \liminf_n \ang{ v, \pi_n} \geq \liminf_n \ang{ \lsc v, \pi_n} \geq \ang{\lsc v, \pi},
\]
where the first inequality follows from the pointwise bound $v \geq \lsc v$, and the second inequality follows from the portmanteau theorem. Therefore, it suffices to construct a particular sequence $(\pi_n)$ converging to $\pi$ such that  $\limsup_n \ang{ v, \pi_n} \leq \ang{\lsc v, \pi}$. 

Fix a compatible metric on $\Th$. Choose strictly positive sequences $(\d_n)$ and $(\e_n)$ that each converge to $0$. Since $\Th$ is separable, for each $n$ there exist finitely many $\d_n$-diameter open balls $B_{n,j}$ for $j = 1, \ldots, J_n$ such that $\pi (\cup_{j = 1}^{J_n} B_{n,j}) \geq 1 - \e_n$. To simplify notation below, let $B_{n, J_n + 1 }= \Th$. For each $j =1 ,\ldots, J_n + 1$, choose $\th_{n,j} \in B_{n,j}$ such that\footnote{First, choose $\th_{n,j}' \in B_{n,j}$ such that $\lsc v ( \th_{n,j}') \leq \inf_{\th \in B_{n,j}} \lsc v(\th) + \e_n/2$. Since $B_{n,j}$ is open, we can then choose $\th_{n,j} \in B_{n,j}$ such that $v(\th_{n,j}) \leq \lsc v (\th_{n,j}') + \e_n/2$.} 
    \[
        v ( \th_{n,j}) \leq \inf_{\th \in B_{n,j}} \lsc v(\th) + \e_n.
    \]
For each $n$ and each $j = 1, \ldots, J_{n+1}$, let $E_{n,j} = B_{n,j} \setminus \cup_{k=1}^{j - 1} B_{n, k}$. Let
\[
    \pi_n = \sum_{j = 1}^{J_n+1} \pi (E_{n,j}) \d_{\th_{n,j}}.
\]
By construction, 
\begin{equation*}
\begin{aligned}
    \ang{ v, \pi_n} 
    &= \sum_{j=1}^{J_n + 1} \pi \Paren{ E_{n,j} } v(\th_{n,j}) \\
    &\leq \e_n + \sum_{j=1}^{J_n + 1} \Ang{ 1_{E_{n,j}} \lsc v, \pi} \\
    &= \e_n + \ang{\lsc v , \pi}.
\end{aligned}
\end{equation*}
Since $\e_n \to 0$, we conclude that $ \limsup_n \ang{ v, \pi_n}    \leq \ang{ \lsc v, \pi}$. To see that $(\pi_n)$ converges to $\pi$, note that for each $n$, the Prokhorov distance between $\pi_n$ and $\pi$ is at most $\max \{ \d_n, \e_n\}$, and we have $\max \{ \d_n, \e_n\} \to 0$.

\subsection{Proof of Lemma~\ref{res:approximation}}

For this proof, we introduce additional notation. For any measure $\mu$ in $\D(\Th)$ and any $\mu$-integrable function $f \colon \Th \to \R_+$, define the measure $f \mu$ by
\[
    ( f \mu) (A) = \int_{A} f(\th) \de \mu (\th), 
\]
for each Borel set $A$. 

For the proof, it suffices to consider $\e \in (0,1)$. By Prokhorov's theorem \citep[Theorem 5.2, p.~60]{Billingsley1999}, the sequences $(\pi_n)$ and $(\pi_n')$ are both tight. Thus, there exists a compact set $K$ such that for all $n$ we have $\pi_n (K) \geq 1 - \e$ and $\pi_n'(K) \geq 1 - \e$.\footnote{Since $(\pi_n)$ is tight, there exists a compact subset $K_1$ such that $\pi_n (K_1) \geq 1- \e$ for all $n$. Similarly, since $(\pi_n')$ is tight, there exists a compact subset $K_2$ such that $\pi_n' ( K_2) \geq 1 - \e$ for all $n$. Let $K = K_1 \cup K_2$. The set $K$ is compact and satisfies the desired inequalities.} The function $H$, being continuous, must achieve a maximum over the compact set $K$. Let $L = 1 +  \max_{\th \in K} H(\th)$. Let $C = \{ \th \in \Th: H (\th) \geq L\}$. Define $b \colon \Th \to [0,1]$ by $b(\th)= \min \{ 1, (L - H(\th))_+\}$. The function $b$ is continuous since $H$ is continuous. Note that $b$ equals $1$ on $K$ and equals $0$ on $C$. For each $n$, let
\[
    x_n = \ang{ b, \pi_n} \quad \text{and} \quad x_n' = \ang{ b, \pi_n'}.
\]
For each $n$, we have $x_n \geq \pi_n (K) \geq 1 - \e$ and $x_n' \geq \pi_n' (K) \geq 1- \e$. Let $x = \ang {b, \pi}$. For each $n$, define the nonnegative measure $\rho_n$ by 
\[
  \rho_n = (x_n'/x_n) b \pi_n  + (1 - b) \pi_n'.
\]
Note that $\rho_n (\Th) = x_n' + (1 - x_n') =  1$. Since $(\pi_n)$ and $(\pi_n')$ both weakly converge to $\pi$, we have $x_n'/x_n \to x/x = 1$. Hence, $(\rho_n)$ weakly converges to $\pi$. 

It remains to check the three properties. 
\begin{enumerate}[label = 
(\roman*)]
\item For each $n$, note that $\min \{ x_n'/x_n, 1\} b \pi_n$ is a nonnegative measure that is setwise smaller than both $\pi_n$ and $\rho_n$. Since $(\min \{ x_n'/x_n, 1\} b \pi_n) (\Th) = \min \{ x_n', x_n\}$, it follows that
\[
    \| \rho_n - \pi_n \|_{\mathrm{TV}} \leq 1 - \min \{ x_n', x_n\} \leq \e.
\]

\item For each $n$ and any Borel subset $A$ of $C$, we have $\rho_n (A) = \pi_n'(A)$, so $\supp (\rho_n - \pi_n') \subseteq \overline{\Th \setminus C}$. The function $H$ is bounded above by $L$ on $\Th \setminus C$, and hence on the closure $\overline{\Th \setminus C}$ by continuity. Therefore, the nonnegative function $H$ is bounded above by $L$ on $\cup_n \supp (\rho_n - \pi_n')$. 

\item Let $h \colon \Th \to \R$ be a continuous function satisfying $|h| \leq H$. By (ii), we know that 
for each $n$, the integral $\ang{ h, \rho_n - \pi_n'}$ is well-defined and finite. For each $n$, we have 
\[
    \rho_n - \pi_n' =  (x_n'/x_n) b \pi_n - b \pi_n'.
\]
Since $|h| \leq H$, the function $hb$ is bounded and continuous. Therefore,
\[
    \ang{ h,     \rho_n - \pi_n' } = (x_n'/x_n) \ang{ hb, \pi_n} - \ang{ h b, \pi_n'} \to \ang{ hb, \pi} - \ang{hb, \pi} = 0,
\]
where we have used the fact that $x_n'/x_n \to 1$.

\end{enumerate}

\subsection{Proof of Lemma~\ref{res:concentration}}

Fix a prior $\pi$ in $\D(\Th)$. We first construct a dense Borel subset $N$ of $\Th$ with $\pi(N) = 0$. Let $\Th_0$ consist of all points in $\Th$ with positive $\pi$-measure. Since $\pi(\Th) = 1 < \infty$, the set $\Th_0$ is countable. For each $\th$ in $\Th$, the complement $\Th \setminus \{ \th \}$ is open and dense (since $\th$ cannot be an isolated point because $\Th$ is perfect). By the Baire category theorem, the set $\Th \setminus \Th_0 = \cap_{\th \in \Th_0} (\Th \setminus \{\th\})$ is dense as well. Since $\Th$ is separable, there exists a countable subset $N$ of $\Th \setminus \Th_0$ that is dense in $\Th$. Thus, $\pi (N) = 0$. Let $A = \Th \setminus N$.

Let $(\th_j)$ be an enumeration of $N$. Select a bounded, compatible metric $d$. For each $n$, we can partition $\Th$ as $\cup_{j=1}^{\infty} E_{n,j}$, where $E_{n,j} = B( \th_j, 1/n) \setminus \cup_{k=1}^{j-1} B ( \th_k, 1/n)$ for all $j$. For each $n$, define the probability measure
\[
    \pi_n = \sum_{j=1}^{\infty} \pi (E_{n,j}) \d_{\th_j}.
\]
Let $W$ be the Wasserstein metric on $\D(\Th)$ associated with $d$. By construction, $W ( \pi_n, \pi) \leq 1/n$, so the sequence $(\pi_n)$ converges to $\pi$ in the Wasserstein metric, and hence weakly, by \citet[Corollary 6.13, p.~97]{Villani2009}.

\end{document}